\documentclass[pre,twocolumn,floatfix,showpacs,amsmath,amssymb]{revtex4}
\usepackage{graphicx}
\begin{document}
\title{Statistics of Pressure Fluctuations in Decaying, Isotropic Turbulence}
\author{Chirag Kalelkar}
\email{kalelkar@physics.iisc.ernet.in}
\affiliation{Centre for Condensed Matter Theory, Department of Physics, 
Indian Institute of Science, Bangalore 560012, India.}
\begin{abstract}
We present results from a systematic direct-numerical simulation study 
of pressure fluctuations in an unforced, incompressible, homogeneous, and 
isotropic, three-dimensional turbulent fluid. At cascade completion, 
isosurfaces of low pressure are found to be organised as slender filaments, 
whereas the predominant isostructures appear sheet-like. We exhibit 
several new results, including plots of probability distributions of the 
spatial pressure-difference, the pressure-gradient norm, and the eigenvalues 
of the pressure-hessian tensor. Plots of the temporal evolution of the mean 
pressure-gradient norm, and the mean eigenvalues of the pressure-hessian 
tensor are also exhibited. We find the statistically preferred orientations 
between the eigenvectors of the pressure-hessian tensor, the 
pressure-gradient, the eigenvectors of the strain-rate tensor, the 
vorticity, and the velocity. Statistical 
properties of the non-local part of the pressure-hessian tensor are also 
exhibited, for the first time. We present numerical tests (in the viscous 
case) of some conjectures of Ohkitani [Phys. Fluids A {\bf 5}, 2570 (1993)] 
and Ohkitani and Kishiba [Phys. Fluids {\bf 7}, 411 (1995)] concerning the 
pressure-hessian and the strain-rate tensors, for the unforced, 
incompressible, three-dimensional Euler equations.
\end{abstract}
\maketitle
\section{Introduction}
The pressure at a point in an incompressible fluid is a non-local functional 
of the velocity field, and inherently difficult to measure in laboratory 
experiments. However, high-resolution direct-numerical simulations (DNS) 
provide accurate statistics of the pressure field. Pressure fluctuations have 
been extensively studied in both numerical 
studies\cite{Pumir,Cao,Gotoh,Ishihara,Ashurst,She} and laboratory 
experiments\cite{Fauve,Abry,Douady,Villermaux} of {\it statistically steady}, 
homogeneous, and isotropic turbulent flows. High-resolution numerical 
studies\cite{Pumir,Cao,Gotoh,Ishihara} show, 
that the pressure spectrum exhibits a wavenumber range with power-law 
scaling, in accordance with predictions from a phenomenological theory 
due to Kolmogorov\cite{Monin}. 
The pressure probability distribution is widely accepted to be negatively 
skewed, and to exhibit an exponential low-pressure tail\cite{Fauve,Abry}. 
Regions of low pressure are found to be organised as slender filamentary 
structures in both numerical studies\cite{Cao} and laboratory 
experiments\cite{Douady,Villermaux}.

The `canonical' isotropic turbulent flow is a {\it decaying} turbulent flow 
behind a grid\cite{Monin}. The study of decaying turbulence is important 
since the results are uninfluenced by statistics of the forcing and 
directly reflect effects of the nonlinear terms in the Navier-Stokes equations 
(see Eqs. (\ref{eq:nse}) below). In contrast to statistically steady 
turbulence, systematic numerical studies of pressure fluctuations within 
the context of decaying, homogeneous, and isotropic turbulence are 
extremely scarce. The only 
available work is a low-resolution DNS study due to Schumann and 
Patterson\cite{Schumann} who exhibited plots of the 
root-mean-square pressure fluctuations as a function of the time, and the 
isosurfaces of low pressure. The low-pressure isosurfaces in this 
study\cite{Schumann} were shown to be organised as `cloud'-like structures, in 
contrast to the slender filaments seen in DNS studies\cite{Cao} of 
statistically steady turbulence at higher resolutions. The pressure-hessian 
tensor and the pressure-gradient were not studied in this work\cite{Schumann}. 
In both statistically steady and decaying turbulence, a comprehensive study of 
possible alignments between vectors of interest in a turbulent flow, 
namely, the eigenvectors of the pressure-hessian tensor, the 
pressure-gradient, the eigenvectors of the strain-rate tensor, the 
vorticity, and the velocity, is entirely lacking.\\
In this paper, we present results from a systematic numerical study of 
the pressure, the pressure-gradient, and the pressure-hessian tensor in 
an unforced, incompressible, homogeneous, and isotropic turbulent fluid. We 
exhibit several new results, including plots of the probability 
distributions of the spatial pressure-difference, the pressure-gradient 
norm, and the eigenvalues of the pressure-hessian tensor, temporal 
evolution of the mean pressure-hessian eigenvalues and of the mean 
pressure-gradient norm, as well as isosurfaces of the pressure and 
the pressure-gradient norm, at cascade 
completion. Statistical properties of the non-local part of the 
pressure-hessian tensor are also exhibited, for the first time. We construct 
the general alignment picture between the eigenvectors of the pressure-hessian 
tensor, the pressure-gradient, the eigenvectors of the strain-rate tensor, the 
vorticity, and the velocity. Ohkitani\cite{Ohkitani1} and Ohkitani and 
Kishiba\cite{Ohkitani2} have derived several interesting results for the 
unforced, incompressible, three-dimensional, inviscid Navier-Stokes 
equations (the Euler equations) concerning the pressure-hessian and the 
strain-rate tensors. We exhibit numerical tests of some conjectures for the 
Navier-Stokes case. 

The unforced Navier-Stokes equations are
\begin{eqnarray}
&&\frac{\partial{\bf v}}{\partial{t}}+({\bf v\cdot\nabla}){\bf v}=
-\frac{\nabla p}{\rho}+\nu{\nabla^2\bf v},
\label{eq:nse}
\end{eqnarray}
where $\nu$ is the kinematic viscosity, $\rho$ is the (constant) density, and 
$p$ is the pressure. On taking the divergence of Eqs. (\ref{eq:nse}), and 
enforcing incompressibility $\nabla\cdot{\bf v}=0$, the pressure is determined 
by
\begin{equation}
\nabla^2p=\Big(\frac{1}{2}\omega^2-s^2\Big)\rho,
\label{eq:peqn}
\end{equation}
where the enstrophy $\omega^2=\omega_i\omega_i$, 
$\omega_i\equiv\epsilon_{ijk}\partial_jv_k$ is the 
vorticity ($\epsilon_{ijk}$ is the Levi-Civita tensor),  $i,j,k=1,2,3$, with 
a summation implicit over repeated indices, and $s^2=S_{ij}S_{ij}$, 
$S_{ij}\equiv1/2(\partial_jv_i+\partial_iv_j)$ is the strain-rate tensor. 
\section{Numerical Method}
We use a pseudospectral method\cite{Dhar} to solve Eqs. (\ref{eq:nse}) 
numerically, in a cubical box of side $2\pi$ with periodic boundary conditions 
and $224^3$ Fourier modes. In this paper, we do not address issues pertaining 
to the scaling of higher-order structure functions of the pressure-difference 
(or pressure-velocity correlations) or investigate dissipation-scale 
properties, and believe that our spectral resolution is 
adequate for the types of studies that we have carried out (barring the 
pressure spectrum, see below). We have checked that our results are unaffected 
by resolution considerations, by comparing with results from a $128^3$ DNS 
study with identical (initial) Reynolds number. For the temporal evolution, we 
use an Adams-Bashforth scheme 
(step size $\delta t=10^{-3}$) with double-precision arithmetic and set 
$\rho=1$, $\nu=10^{-5}$. We include a hyperviscous term of the form 
$\nu_h\nabla^4{\bf v}$ in Eqs. (\ref{eq:nse}), with $\nu_h=10^{-6}$ and have 
explicitly checked that our results are unaffected by the inclusion of 
hyperviscosity. We note that Borue and Orszag\cite{Borue} have carried out 
a $256^3$ DNS study of decaying, isotropic turbulence with hyperviscosity, 
and they conclude that ``...inertial-range dynamics may be independent of the 
particular mechanism of small-scale dissipation..." (p. R$859$, Ref. 
\cite{Borue}). The initial velocity field is taken to be 
${\bf v}({\bf k},t_0)\sim k^2e^{-k^2}e^{i\theta_{\bf k}}$ ($k=|{\bf k}|$ is 
the wavenumber), with $\theta_{\bf k}$ independent random variables 
distributed uniformly between $0$ and $2\pi$. This corresponds to an 
initial kinetic-energy spectrum $E(k,t_0)\sim k^4e^{-2k^2}$ (with 
$E(k,t)\equiv|{\bf v}({\bf k},t)|^2$ the one-dimensional spectrum), which is a 
convenient choice that develops a cascade to large wavenumbers (see below). 
We measure time in units of the initial `box-size' time 
$\tau_0\equiv2\pi/v_{rms}^0$ (here  $\tau_0$ equals $4.02$), 
$v^0_{rms}\equiv[\langle\sum_{{\bf k}}|{\bf v}({\bf k},t_0)|^2\rangle]^{1/2}$ 
is the root-mean-square value of the initial velocity, with the dimensionless 
time $\tau\equiv t/\tau_0$ ($t$ is the product of the number of steps and 
$\delta t$). We define $Re_0\equiv2\pi v^0_{rms}/\nu$ to be the value of the 
initial `box-size' Reynolds number (here $Re_0$ equals $982464$). Our 
results are obtained for times $t_0\le t<<t_*$, where 
$t_*$ is the time at which the (growing) integral scale 
$L(t)\equiv\langle(\sum_{\bf k}|{\bf v}({\bf k},t)|^2/k)/\sum_{\bf k}
|{\bf v}({\bf k},t)|^2\rangle$ becomes of the order of the linear size of the 
simulation box. For times $t\gtrsim t_*$, finite-size effects which might 
well be non-universal, modify the numerical results, and are not considered 
here.

\begin{figure}
\includegraphics[height=2.2in]{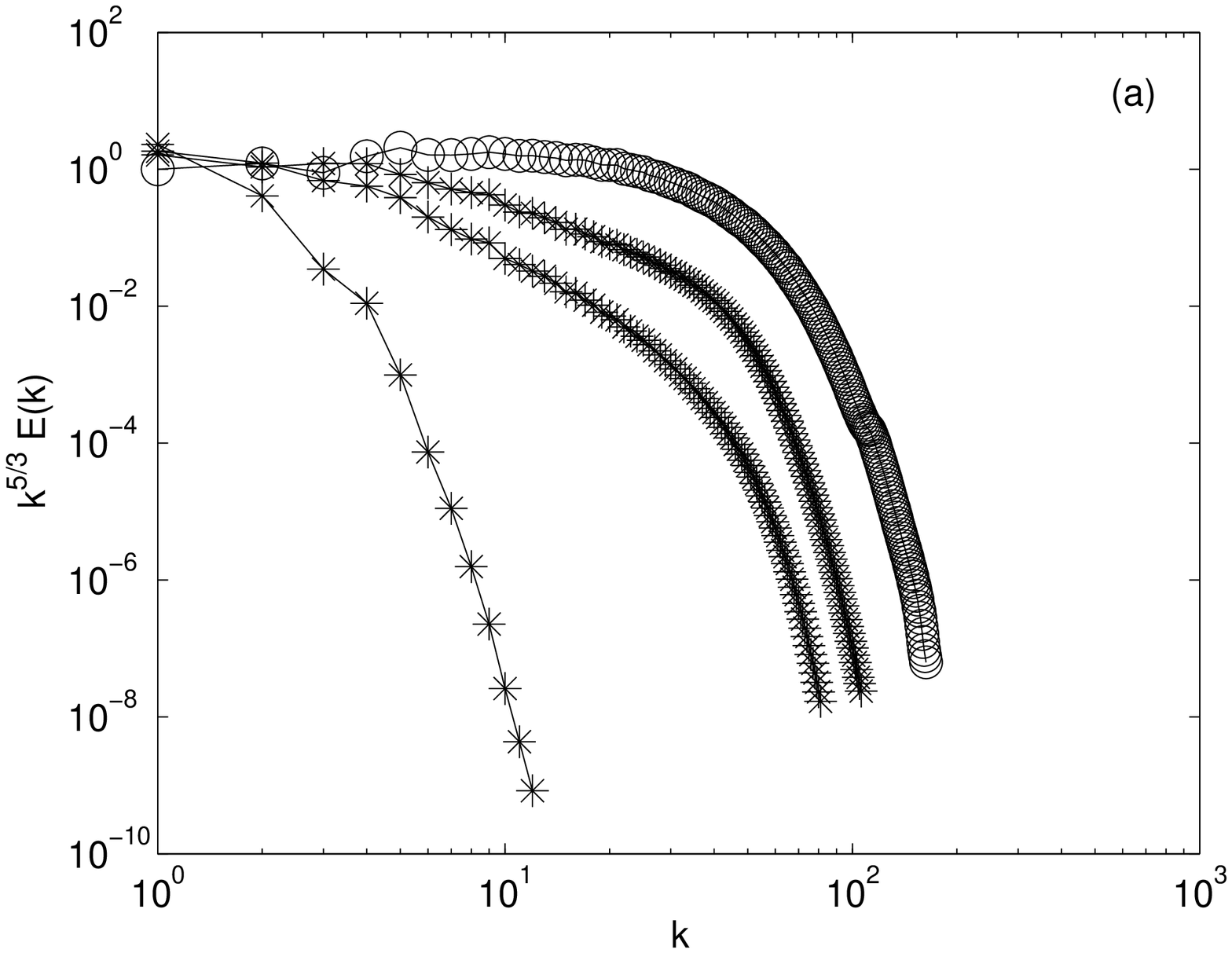}
\includegraphics[height=2.2in]{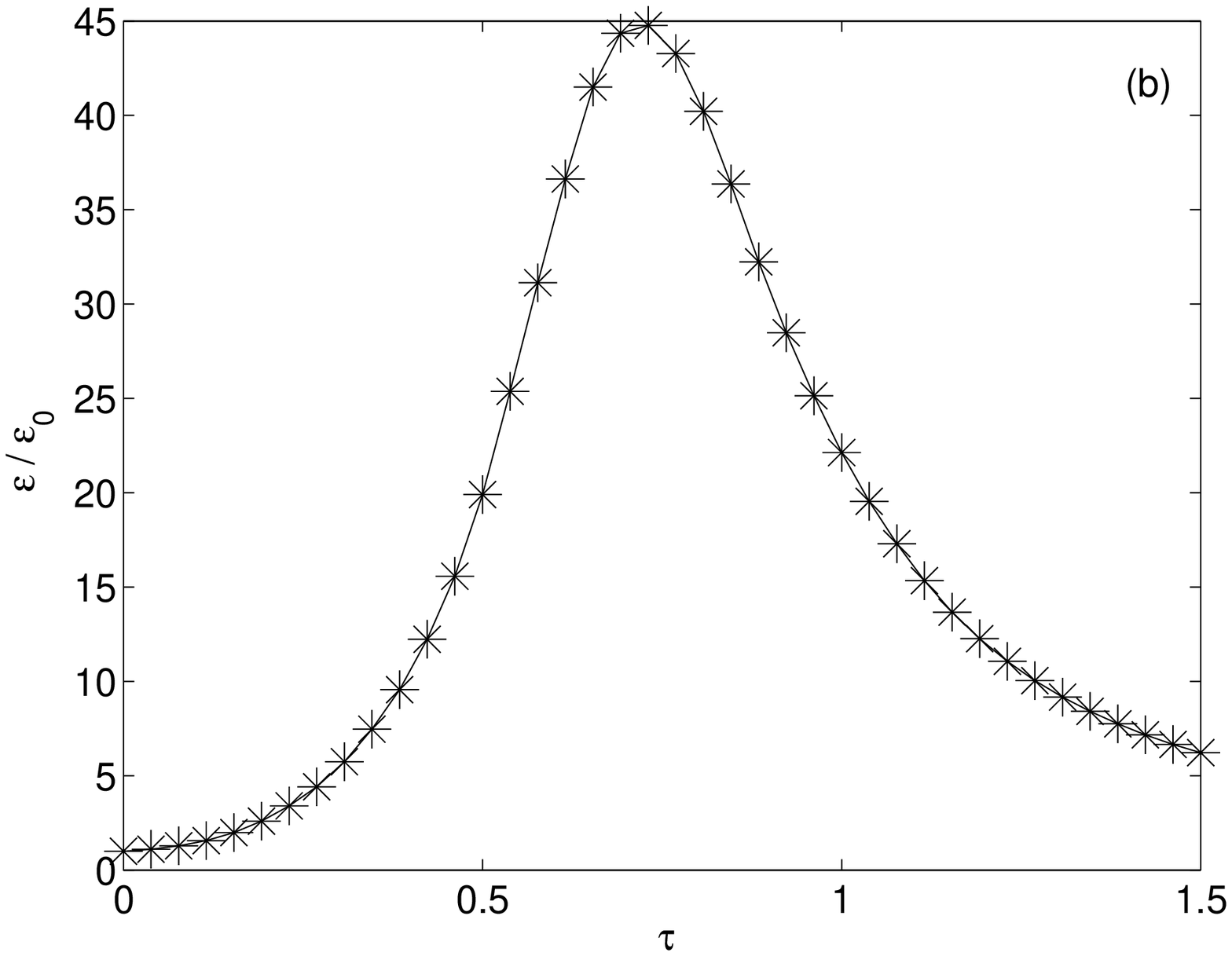}
\caption{\label{fig:prelim}(a) Log-log plot of the temporal evolution of the 
scaled kinetic energy spectrum $k^{5/3}E(k,\tau)$ as a function of the 
wavenumber $k$ at temporal separations of $\tau=0.24$. The plot with open 
circles is calculated at cascade completion, at dimensionless time 
$\tau=\tau_c=0.71$.\\(b) Plot of the normalized kinetic 
energy-dissipation rate $\epsilon(\tau)/\epsilon_0$ as a function of the 
dimensionless time $\tau$.}
\end{figure}

In Fig. \ref{fig:prelim}, we show some preliminary results that serve as a 
check of our numerical method and parameter values (which were chosen 
to ensure linear stability of the numerical scheme). Figure \ref{fig:prelim}(a) 
shows on a log-log plot, the scaled kinetic energy spectrum $k^{5/3}E(k,\tau)$ 
as a function of the wavenumber $k$. On starting with the spectrum specified 
above, a cascade of energy is seen to large wavenumbers. The plots are 
equispaced in time with a temporal separation of $\tau=0.24$. The plot with 
open circles is calculated at cascade completion at the dimensionless time 
$\tau=\tau_c=0.71$, and shows a wavenumber range (for 
$1\lesssim k\lesssim10$) that exhibits the well-known $-5/3$ 
power-law\cite{Monin,Yamamoto,Kalelkar}. Upon cascade completion, the shape of 
the energy spectrum does not change appreciably (except at large 
wavenumbers where it 
falls), but the kinetic energy decays monotonically. In Fig. 
\ref{fig:prelim}(b), we plot the normalized kinetic energy-dissipation rate 
$\epsilon(\tau)/\epsilon_0$ [$\epsilon(t)\equiv\sum_{\bf k}k^2|
{\bf v}({\bf k},t)|^2$] as a function of the dimensionless time $\tau$. The 
kinetic energy-dissipation rate peaks\cite{Yamamoto,Kalelkar} at 
$\tau=\tau_c$, corresponding to cascade completion in the 
kinetic energy spectrum, and decreases thereafter. The turbulence may be 
considered as `fully developed' at $\tau=\tau_c$ and our spatial results 
(see below) will be calculated at this instant of time.
\section{Numerical Results}
\subsection{Pressure}
\begin{figure}
\includegraphics[height=1.9in]{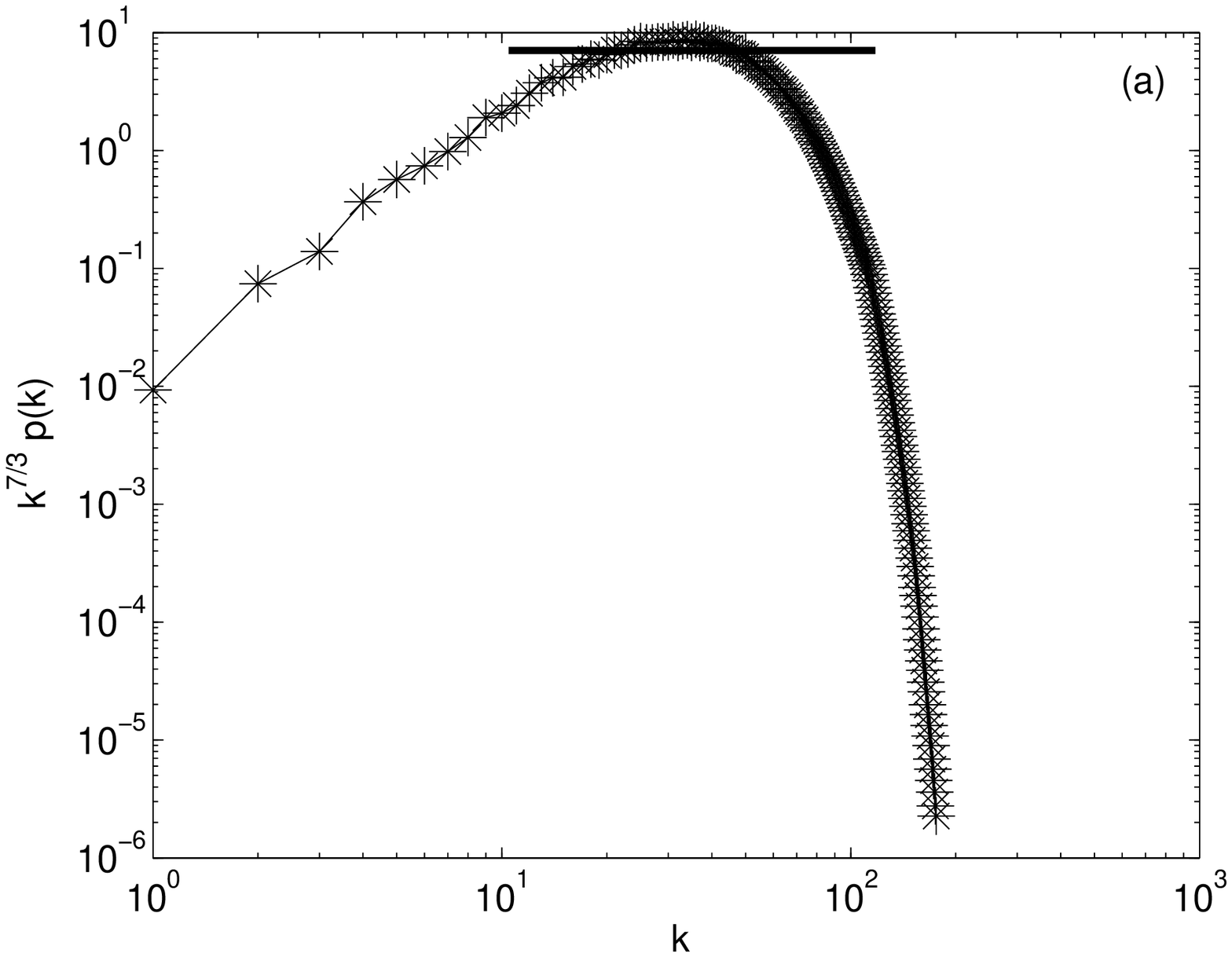}
\includegraphics[height=1.9in]{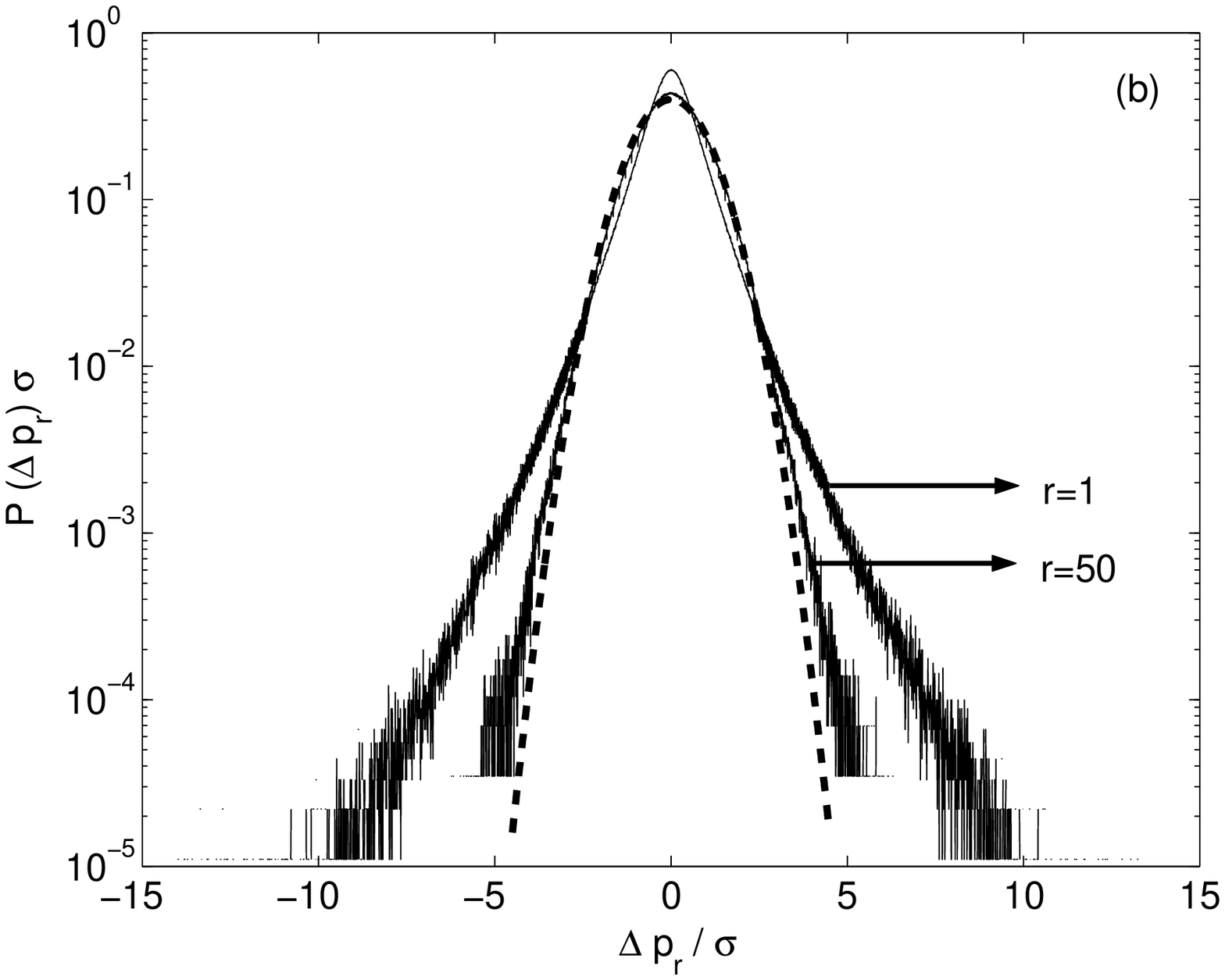}
\caption{\label{fig:pspec}(a) Log-log plot of the scaled 
pressure spectrum $k^{7/3}p(k,\tau)$ as a function of the wavenumber $k$, at 
cascade completion. The horizontal line is drawn for 
reference.\\(b) Semilog plot of the normalized probability distribution 
${\cal P}(\Delta p_r)$ ($\sigma$ is the standard deviation) of the 
pressure-difference $\Delta p_r$ at cascade completion, for grid-spacing 
values $r=1,50$. The dashed-line curve is a normalized Gaussian distribution 
for comparison.}
\end{figure}

At each grid point, we compute $p$ by Fourier-transforming Eq. 
(\ref{eq:peqn}), solving for the pressure, and inverse Fourier-transforming 
to physical space. Kolmogorov phenomenology predicts that the pressure spectrum 
in statistically steady turbulence exhibits a wavenumber range with the 
power-law scaling $p(k)\sim k^{-7/3}$\cite{Monin}. 
Ishihara, {\it et.al.}\cite{Ishihara} have confirmed the scaling law in a 
$2048^3$ DNS study of statistically steady, homogeneous, and isotropic 
turbulence at a Taylor-scale Reynolds number $Re_{\lambda}=732$, whereas 
Tsuji and Ishihara\cite{Tsuji} have observed the scaling law by measuring 
pressure fluctuations in the centre line of a freely decaying turbulent 
jet. In Fig. \ref{fig:pspec}(a), we plot the scaled pressure 
spectrum $k^{7/3}p(k,\tau)$ 
as a function of the wavenumber $k$, at cascade completion. In order to 
observe Kolmogorov-type scaling in the pressure spectrum over a substantial 
wavenumber range, a considerably higher (initial) Reynolds number is 
required\cite{Nelkin} as compared to the result for the kinetic-energy 
spectrum [cf. plot with open circles in Fig. \ref{fig:prelim}(a)]. The 
spectral resolution of our study is inadequate for the 
purposes of fitting a power-law, and the pressure spectrum is found to 
exhibit a $-7/3$ power-law only in the narrow wavenumber range 
$20\lesssim k\lesssim50$.

In Fig. \ref{fig:pspec}(b), we plot the normalized probability distribution 
${\cal P}(\Delta p_r)$ of the pressure-difference 
$\Delta p_r\equiv p({\bf x}+{\bf r})-p({\bf x})$ at cascade completion, for 
grid-spacing values $r\equiv|{\bf r}|=1,50$. For the large seperation 
$r=50$, ${\cal P}(\Delta p_{50})$ is found to be close to a Gaussian 
distribution [cf. dashed-line curve in Fig. \ref{fig:pspec}(b)] with a 
skewness equal to zero, and a kurtosis equal to $3.49$. For $r=1$, 
stretched-exponential tails which are roughly symmetrically placed about 
$\Delta p_1=0$ are observed, with ${\cal P}(\Delta p_1)$ having a 
kurtosis equal to $7.80$. We note that ${\cal P}(\Delta p_1)$ does not 
exhibit a Gaussian core at small values of $\Delta p_1/\sigma$ ($\sigma$ is 
the standard deviation). Our results 
for ${\cal P}(\Delta p_r)$ resemble those obtained for probability 
distribution of the spatial velocity-difference\cite{Vincent} at large and 
small grid-spacings respectively. Cao, {\it et.al.}\cite{Cao} obtained similar 
results for ${\cal P}(\Delta p_r)$ from a $512^3$ DNS study of statistically 
steady, homogeneous, and isotropic turbulence at $Re_\lambda=218$.

\begin{figure}
\includegraphics[height=1.9in]{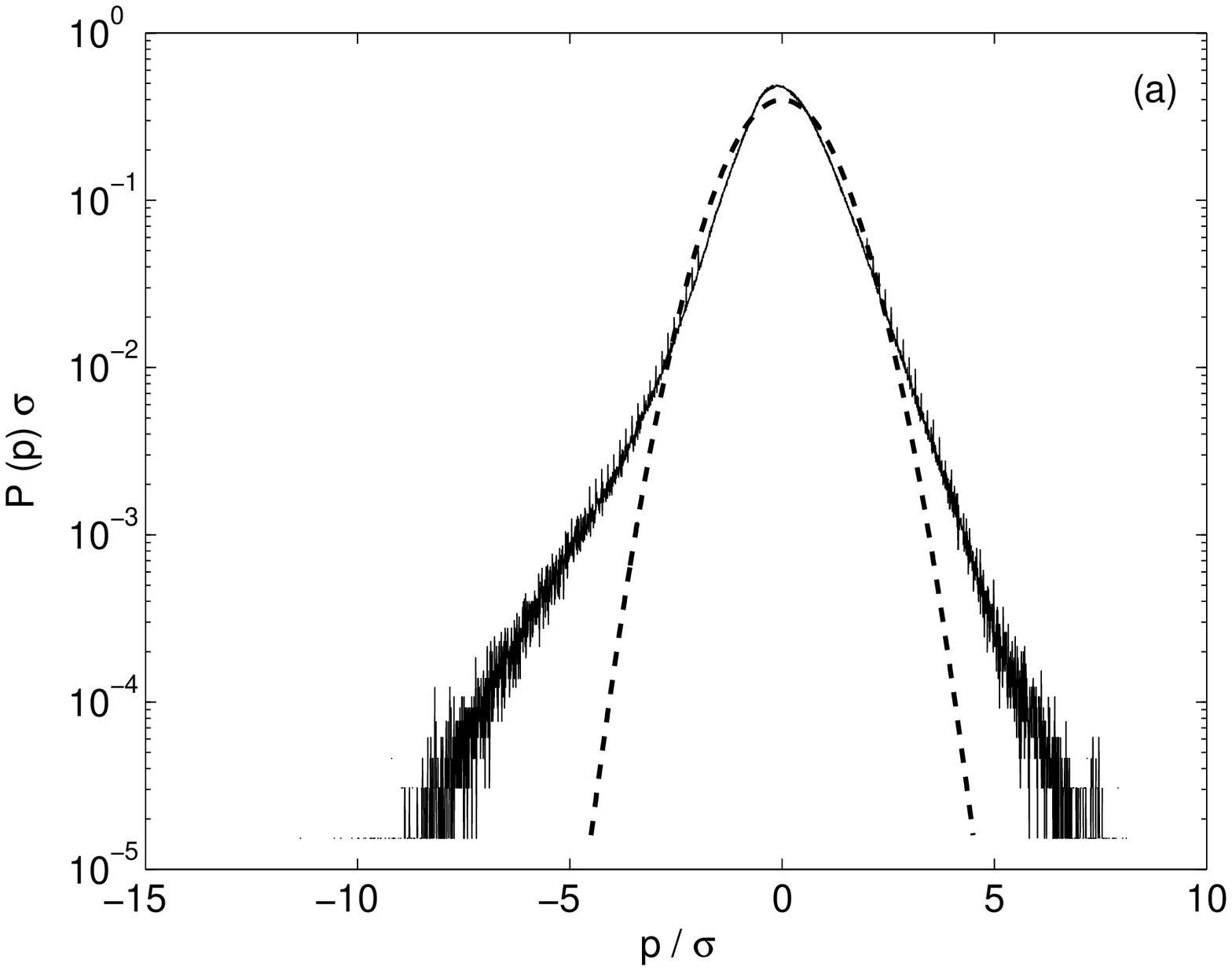}
\includegraphics[height=1.9in]{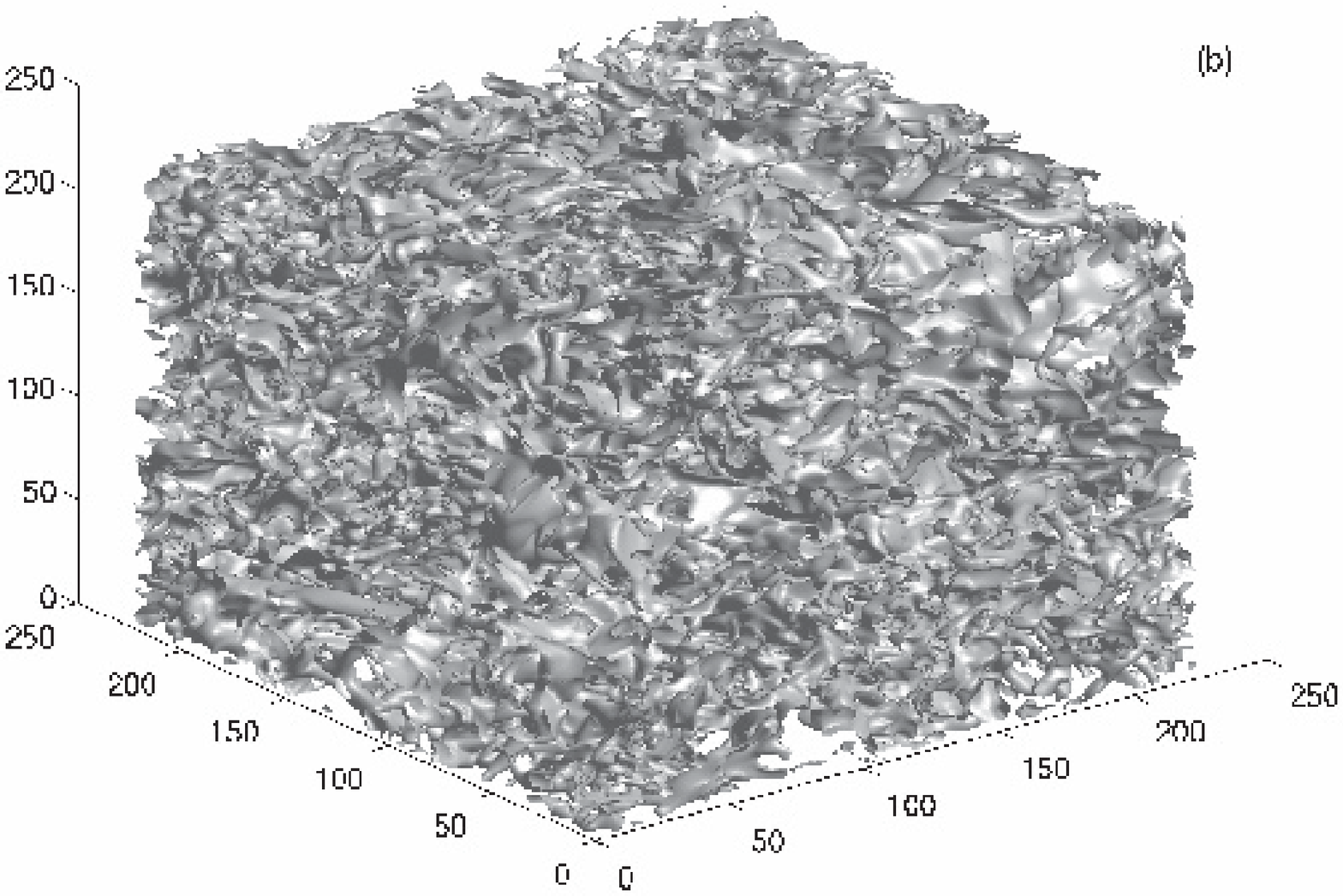}
\caption{\label{fig:pdfp}(a) Semilog plot of the normalized 
probability distribution ${\cal P}(p)$ of the pressure $p$, at cascade 
completion. The dashed-line curve is a normalized Gaussian 
distribution for comparison.\\(b) Plot of iso-$p$ surfaces for the isovalue 
$p=\langle p\rangle$, at cascade completion.}
\end{figure}

\begin{figure}
\includegraphics[height=1.9in]{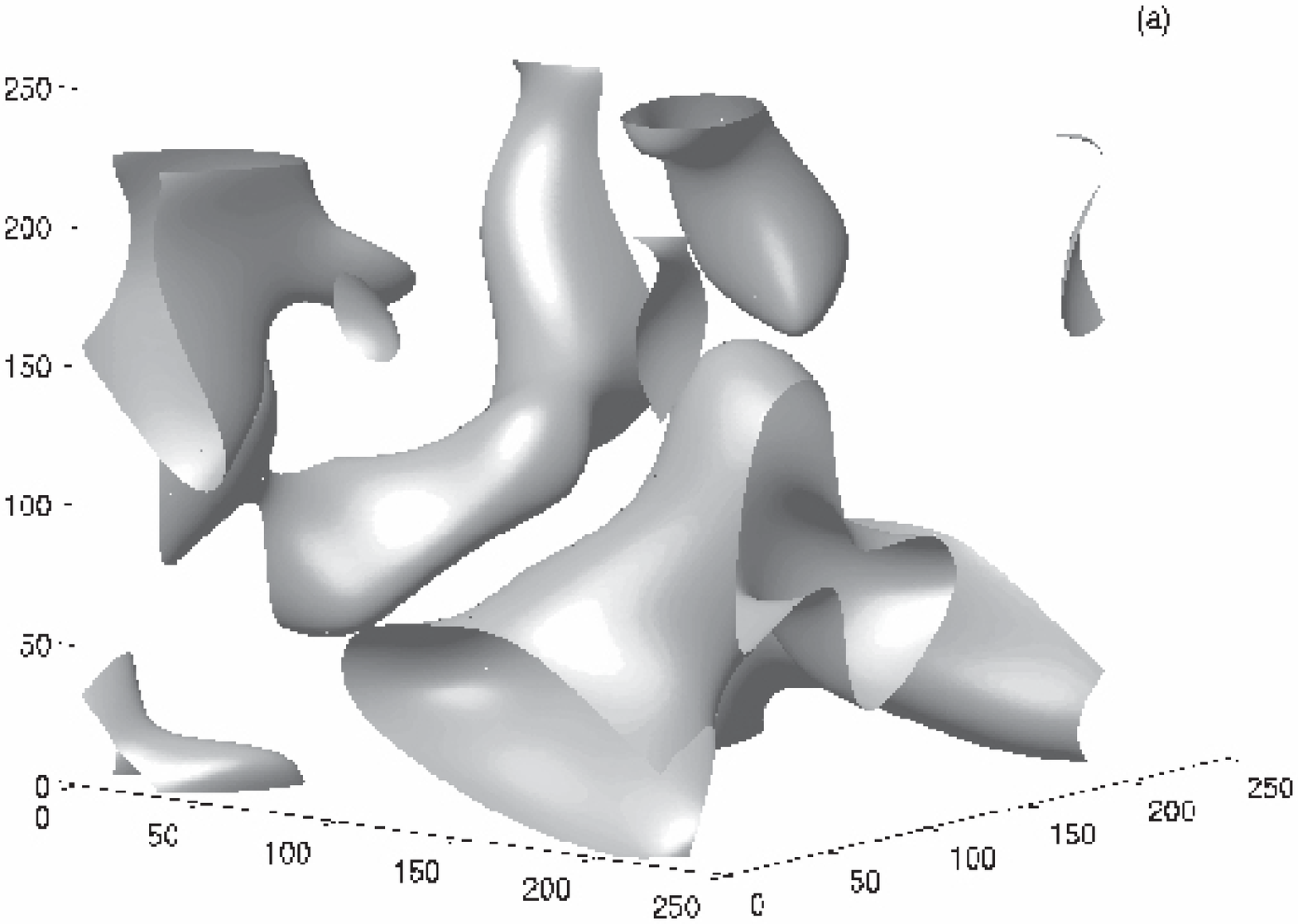}
\includegraphics[height=1.9in]{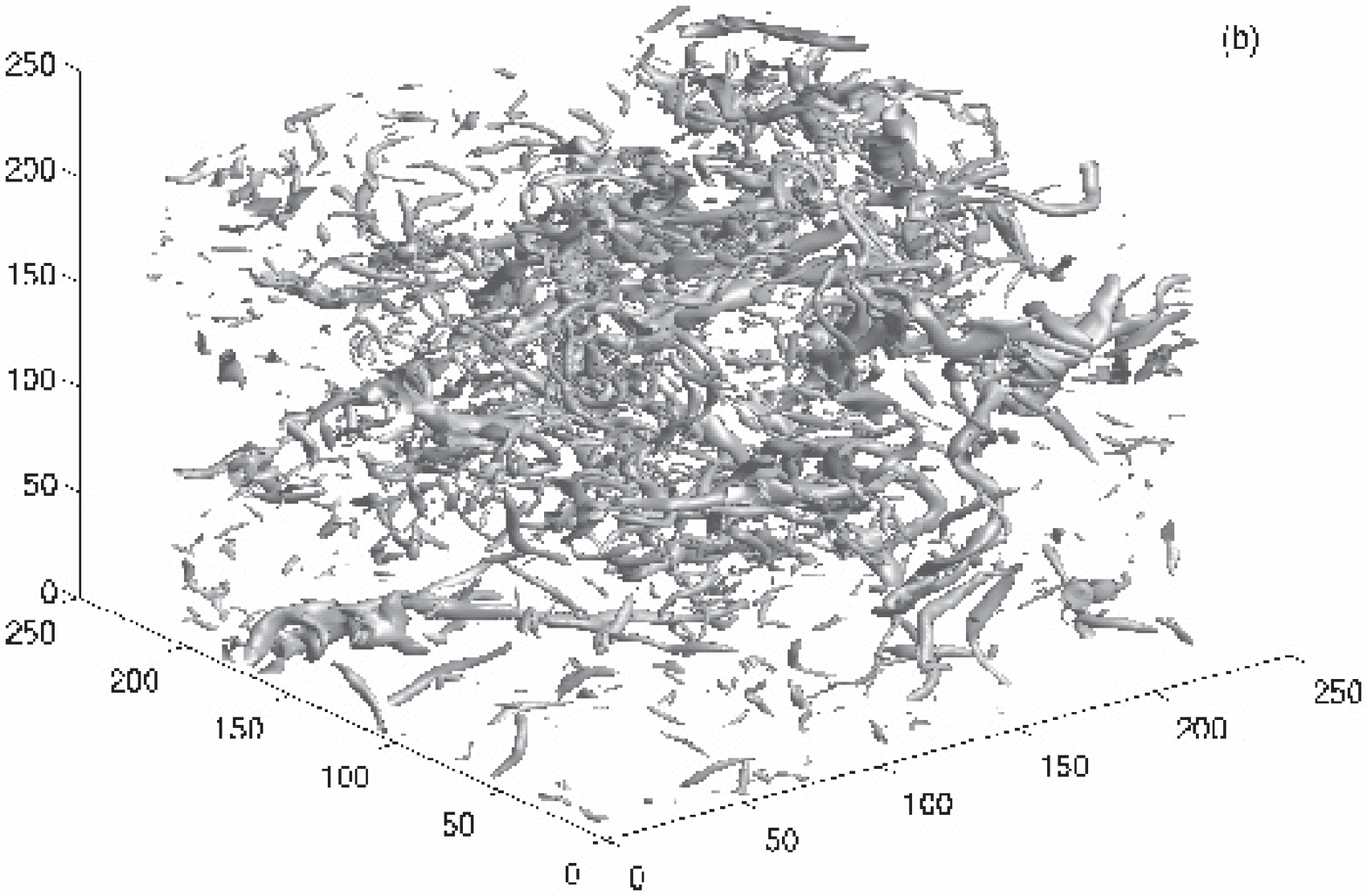}
\caption{\label{fig:isop}(a) Plot of iso-$p$ surfaces for the isovalue 
$p=\langle p\rangle-2\sigma$ at the dimensionless time $\tau<<\tau_c$.\\(b) 
Plot of iso-$p$ surfaces for the isovalue $p=\langle p\rangle-2\sigma$, at 
cascade completion.} 
\end{figure}

In Fig. \ref{fig:pdfp}(a), we plot the normalized probability 
distribution ${\cal P}(p)$ of $p$, at cascade 
completion. The mean pressure $\langle p\rangle$ (angular brackets denote 
a volume average) was found to equal zero, as is expected\cite{Monin} for 
isotropic turbulence, whereas the skewness was found to equal $-0.11$ and 
the kurtosis equalled $5.62$. Holzer and Siggia\cite{Holzer} have shown 
analytically, that the 
probability distribution of the pressure is negatively skewed and has an 
exponential tail, for a {\it Gaussian}\cite{Fng} velocity probability 
distribution. Brachet\cite{Brachet} has found a low-pressure 
exponential tail in an $864^3$ DNS study of decaying, isotropic 
turbulence with Taylor-Green\cite{Taylor} initial 
conditions. The error bars in this study\cite{Brachet} are probably larger 
than those in the data of Fig. \ref{fig:pdfp}(a). Both Pumir\cite{Pumir} and 
Cao, {\it et.al.}\cite{Cao} observed a stretched-exponential tail at low 
pressures, and a roughly Gaussian tail at high pressures, in DNS studies 
of $p$ in statistically steady turbulence. We confirm the result for negative 
pressures in the case of decaying turbulence, where 
the fit ${\cal P}(|p|)\sim e^{-\beta |p|^\alpha}$, $\beta=2.55\pm0.01$, 
$\alpha=1.35\pm0.01$ (error-bars from least-square fits), is observed at 
cascade completion. However, at positive pressures, a stretched-exponential 
tail with $\beta=1.80\pm0.01$, $\alpha=1.52\pm0.02$ (error-bars from 
least-square fits) is observed in our study.\\
In Fig. \ref{fig:pdfp}(b), we plot iso-$p$ surfaces for the 
isovalue $p=\langle p\rangle$ at cascade completion, 
which appear to be crumpled sheet-like structures (found throughout the 
isovalue range $[\langle p\rangle-\sigma,\langle p\rangle+\sigma]$). 
Equation ($\ref{eq:peqn}$) suggests that regions with high vorticity and 
low strain-rates are simultaneously sources of low pressure. Such regions 
with intense vorticity have been observed by Douady, 
{\it et.al.}\cite{Douady} and Villermaux, {\it et.al.}\cite{Villermaux} in 
statistically steady turbulence experiments, by using cavitation as a 
visualization technique, in a liquid seeded with bubbles. Schumann and 
Patterson\cite{Schumann} exhibited plots of low-pressure isosurfaces from 
a $32^3$ DNS study of the unforced, incompressible, three-dimensional 
Navier-Stokes equations, which were shown to be organised as `cloud'-like 
structures. In our study, at early 
times $\tau<<\tau_c$, regions of low pressure (with the 
isovalue $p=\langle p\rangle-2\sigma$) are 
found to be sheet-like [see Fig. \ref{fig:isop}(a)]. At cascade 
completion, low-pressure regions are found to be organised as slender 
filaments [see Fig. \ref{fig:isop}(b)], with diameter of the order of 
the grid spacing, and a contour length that occasionally extends 
nearly to the linear size of the simulation box. We choose to quote 
dimensions of the structures relative to the (fixed) box-size and the 
grid-spacing, since both the Kolmogorov (dissipative) and 
the integral length scales vary in time, in decaying turbulence. Isosurface 
plots of the enstrophy and the squared strain-rates 
show (see Ref. \cite{Cao}) that highly-strained regions occur close to 
regions of intense enstrophy, and Eq. (\ref{eq:peqn}) suggests a 
lack of any well-defined fluid-mechanical structure of the high-pressure 
regions. In our study, iso-$p$ surfaces in the 
range $p>(\langle p\rangle+\sigma)$ were indeed found not to exhibit 
any particular structure at cascade completion. 
\subsection{Pressure-Hessian Tensor}
The pressure-hessian tensor $P_{ij}\equiv\partial_{ij}p$ 
appears\cite{Ohkitani1,Ohkitani2} in the evolution equation for the 
strain-rate tensor $S_{ij}$. At each grid point, 
we compute the eigenvalues $\lambda_{1,P}$, $\lambda_{2,P}$, and 
$\lambda_{3,P}$ (with the convention 
$\lambda_{1,P}\ge\lambda_{2,P}\ge\lambda_{3,P}$) of $P_{ij}$ and the 
corresponding orthonormal eigenvectors $f_1$, $f_2$, and $f_3$. We also 
compute the eigenvalues $\lambda_{1,S}$, $\lambda_{2,S}$, and 
$\lambda_{3,S}$ (with ordering from extensional to compressive strain-rates 
$\lambda_{1,S}\ge\lambda_{2,S}\ge\lambda_{3,S}$) of $S_{ij}$ and the 
corresponding orthonormal eigenvectors $e_1$, $e_2$, and $e_3$.

\begin{figure}
\includegraphics[height=1.9in]{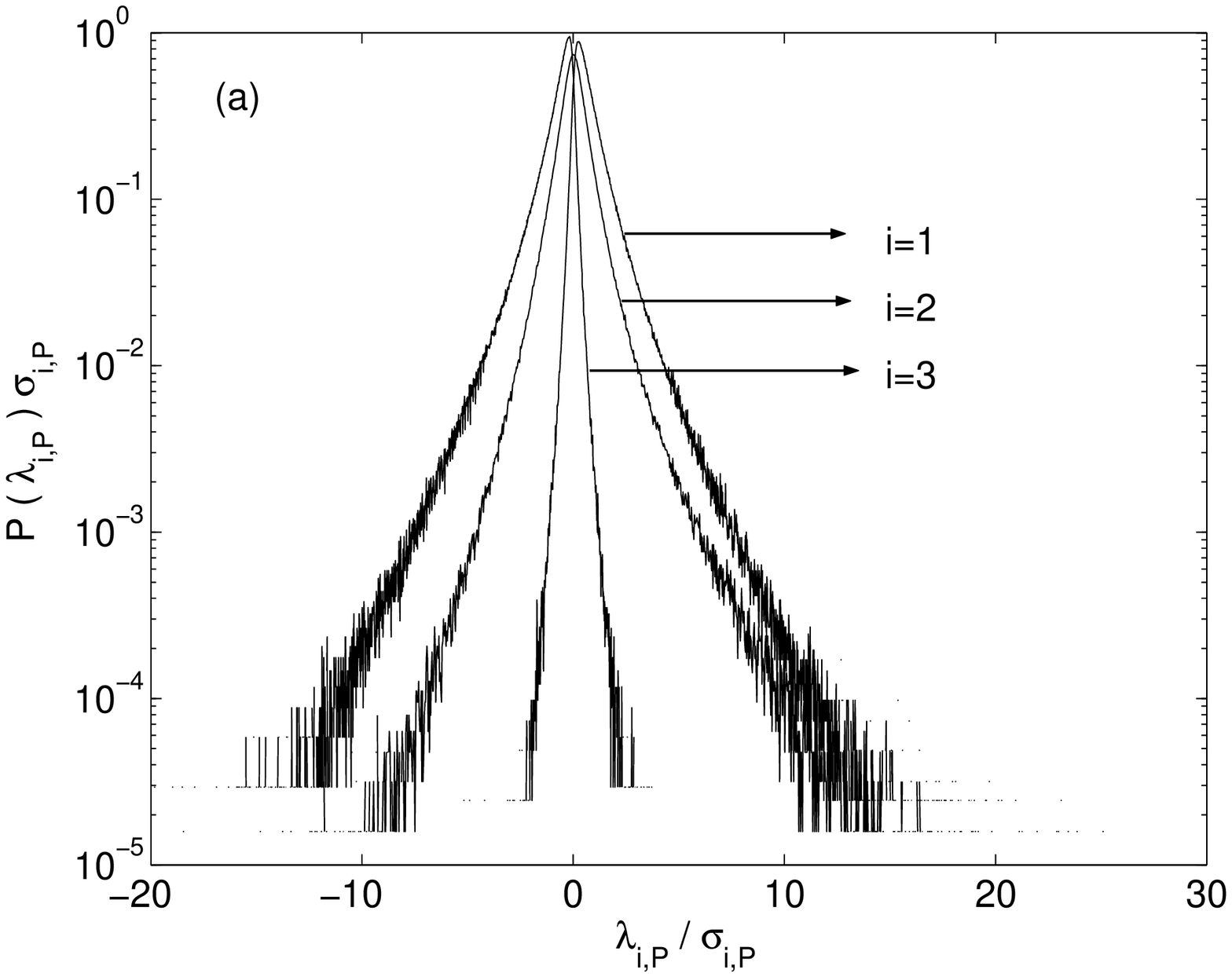}
\includegraphics[height=1.9in]{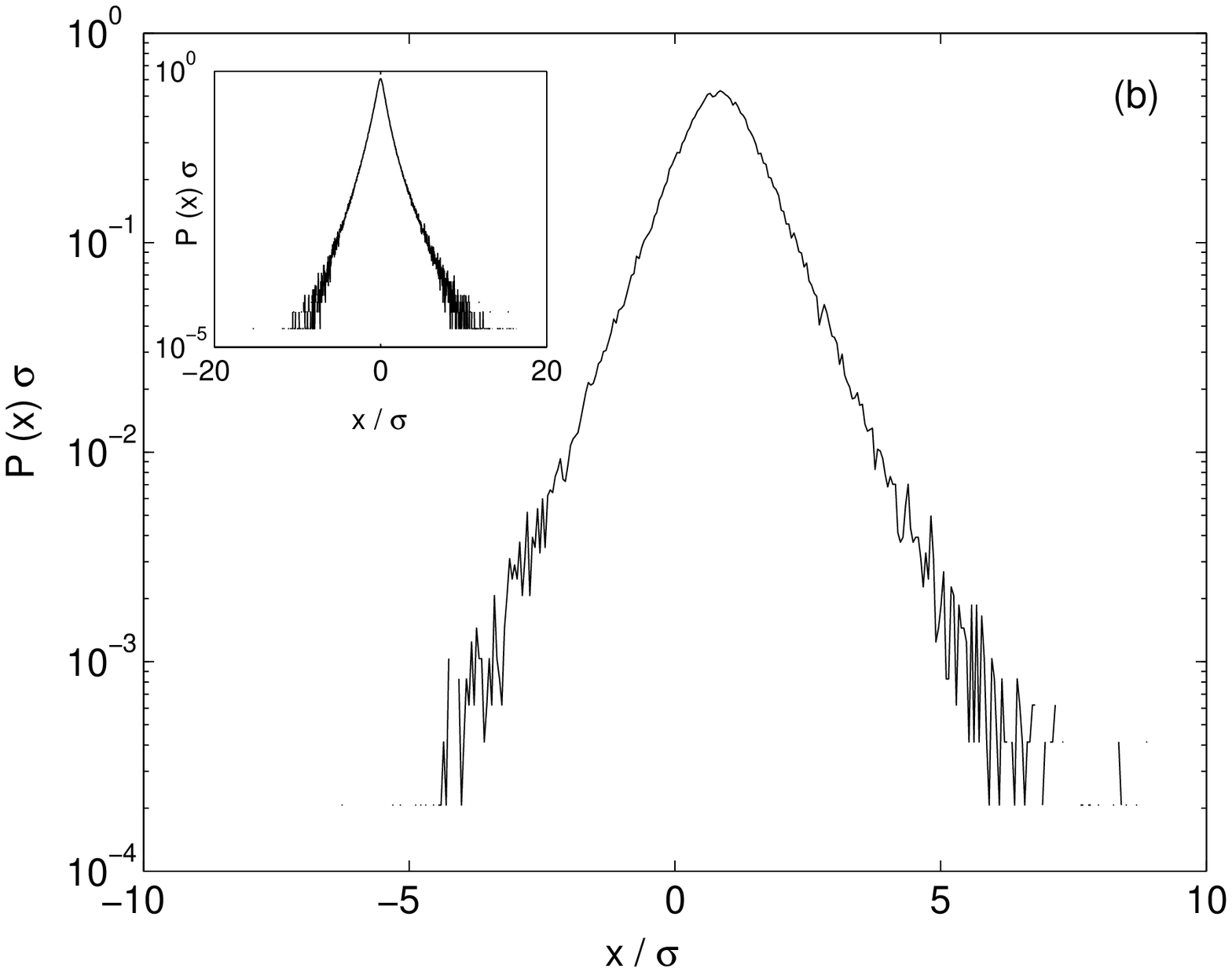}
\includegraphics[height=1.9in]{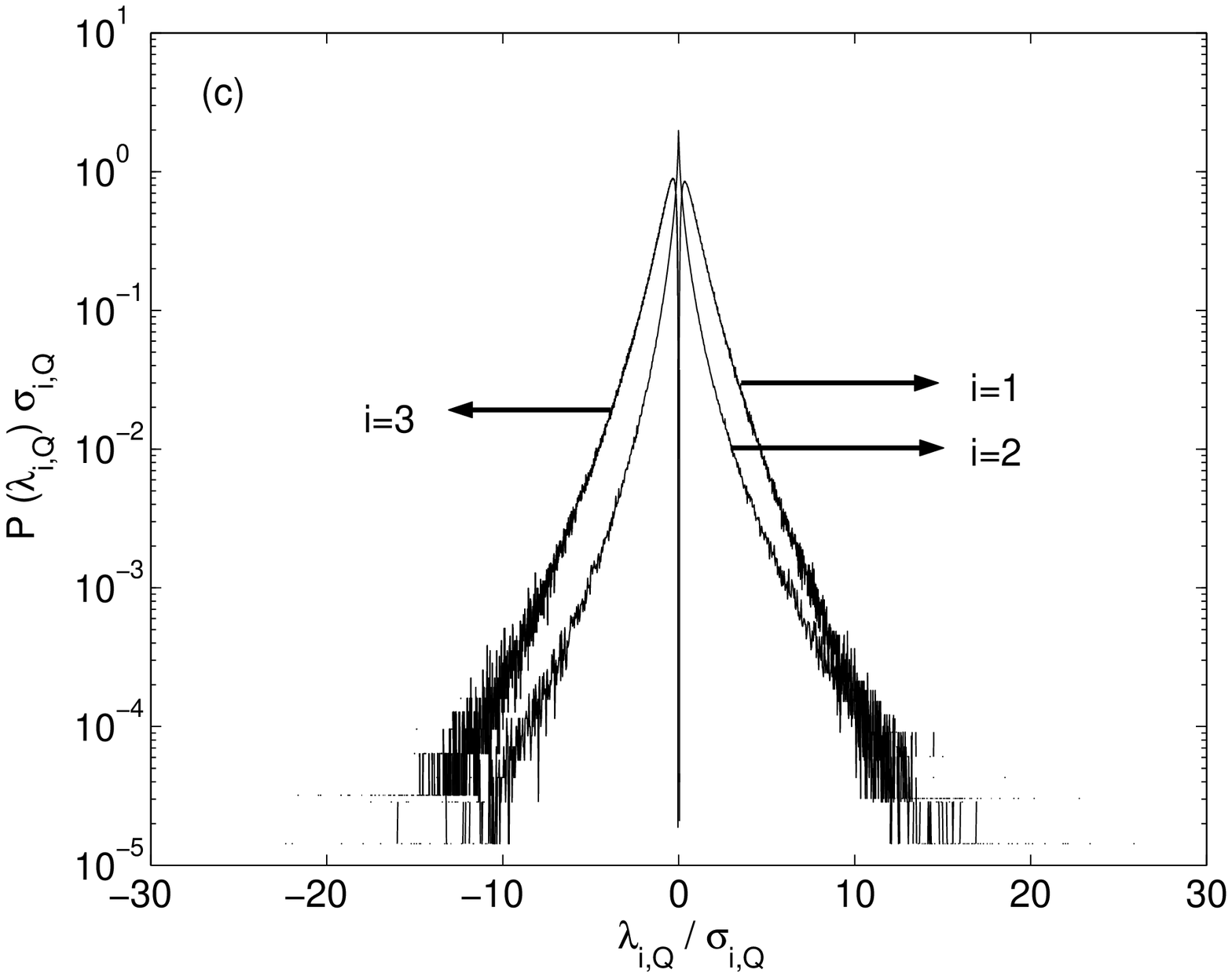}
\caption{\label{fig:peval}(a) Semilog plot of the normalized probability 
distribution ${\cal P}(\lambda_{i,P})$ of the eigenvalues $\lambda_{i,P}$ 
of the pressure-hessian tensor $P_{ij}$, at cascade completion.\\(b) Semilog 
plot of the normalized probability distribution 
${\cal P}[x=Tr(P_{ij})]$ of the trace of 
$P_{ij}$ ($Tr(P_{ij})\equiv\sum_k\lambda_{k,P}$) at cascade completion, 
conditioned on $|\omega|\ge\langle|\omega|\rangle+2\sigma$. The inset is a 
semilog plot of the normalized probability distribution ${\cal P}(x)$ of 
$x=Tr(P_{ij})$, at cascade completion.\\(c) Semilog plot of the normalized 
probability distribution ${\cal P}(\lambda_{i,Q})$ of the eigenvalues 
$\lambda_{i,Q}$ of the tensor $Q_{ij}$ (see the text for the definition), at 
cascade completion.}
\end{figure}

In Fig. \ref{fig:peval}(a), we plot the normalized probability distribution 
${\cal P}(\lambda_{i,P})$ of the eigenvalues $\lambda_{i,P}$, at cascade 
completion. The skewnesses of the eigenvalue 
distributions were found to equal $4.73$, $5.25$, and $-4.13$ for $i=1,2$, 
and $3$ respectively. 

In a constant-density flow, incompressibility requires that 
$\sum_i\lambda_{i,S}=0$, however, there is no such constraint on 
$\lambda_{i,P}$. The inset plot in Fig. \ref{fig:peval}(b) is the normalized 
probability distribution of the trace of $P_{ij}$, ${\cal P}[x=Tr(P_{ij})]$ 
($Tr(P_{ij})\equiv\sum_k\lambda_{k,P}$) at cascade completion, which is 
roughly symmetrically placed about $x=0$. As is well-known in both 
decaying\cite{Kalelkar} and statistically steady\cite{Vincent} turbulence, 
regions of intense vorticity (say, for iso-$|\omega|$ values greater than 
$\langle|\omega|\rangle+2\sigma$) are found to be organised as filamentary 
structures, and are spatially more localised than regions of high strain-rate. 
Equation (\ref{eq:peqn}) suggests that locally, in regions of intense 
enstrophy, $Tr(P_{ij})=\nabla^2p>0$. In Fig. \ref{fig:peval}(b), we plot the 
normalized probability distribution ${\cal P}[Tr(P_{ij})]$ of the trace of 
$P_{ij}$ at cascade completion conditioned on 
$|\omega|\ge\langle|\omega|\rangle+2\sigma$, which is found to have a 
positive mean as expected. 

Ohkitani and Kishiba\cite{Ohkitani2}, have shown that the pressure-hessian 
tensor $P_{ij}$ can be decomposed into the sum of a diagonal tensor 
$\delta_{ij}\nabla^2p/3$ (the `local' term), $\delta_{ij}$ is the Kronecker 
delta, and a symmetric, zero-diagonal tensor $Q_{ij}$\cite{Fnloc} (the 
`non-local' term). The local term, as the name indicates, can be expressed 
purely in terms of the vorticity and the strain-rate at each point of the 
fluid [see Eq. (\ref{eq:peqn})], however, the non-local term can be expressed 
only in terms of an integral over the entire fluid volume. Ohkitani and 
Kishiba\cite{Ohkitani2} have also shown (for Taylor-Green\cite{Taylor} initial 
conditions) that the non-local term contributes significantly to enstrophy 
growth. Since $Q_{ij}=P_{ij}-\delta_{ij}\nabla^2p/3$ has zero 
trace, its eigenvalue $\lambda_{1,Q}>0$, $\lambda_{3,Q}<0$, and the sign of 
$\lambda_{2,Q}$ is indeterminate. In Fig. \ref{fig:peval}(c), we plot the 
normalized probability distribution ${\cal P}(\lambda_{i,Q})$ of the 
eigenvalues $\lambda_{i,Q}$, at cascade completion. We find that 
$\lambda_{2,Q}$ has a positive mean. The statistically preferred ratio of the 
mean eigenvalues $\langle\lambda_{1,Q}\rangle:\langle\lambda_{2,Q}\rangle:\langle\lambda_{3,Q}\rangle$ was found to equal $46:1:-47$, at cascade 
completion\cite{Fneig}. 

\begin{figure}
\includegraphics[height=1.9in]{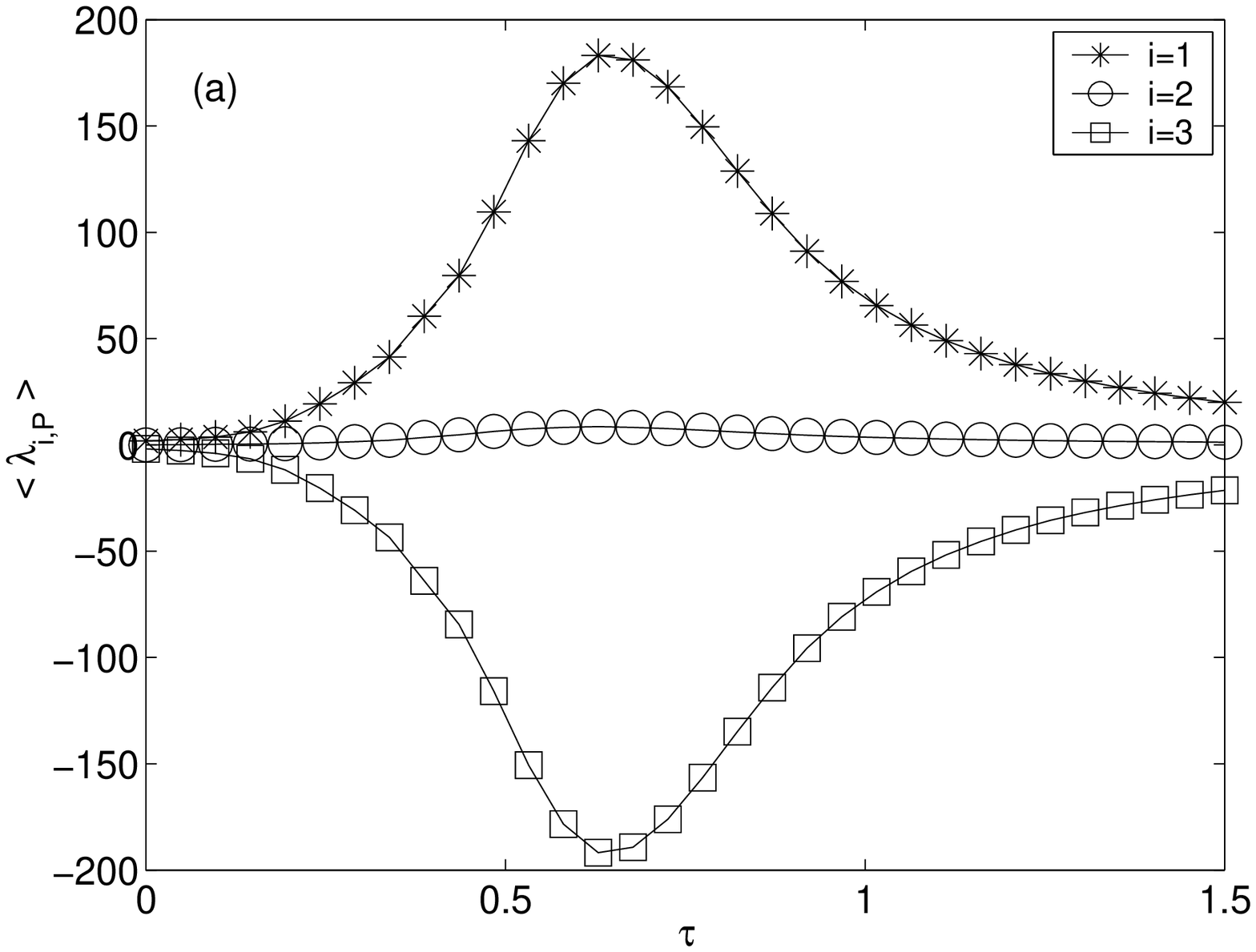}
\includegraphics[height=1.9in]{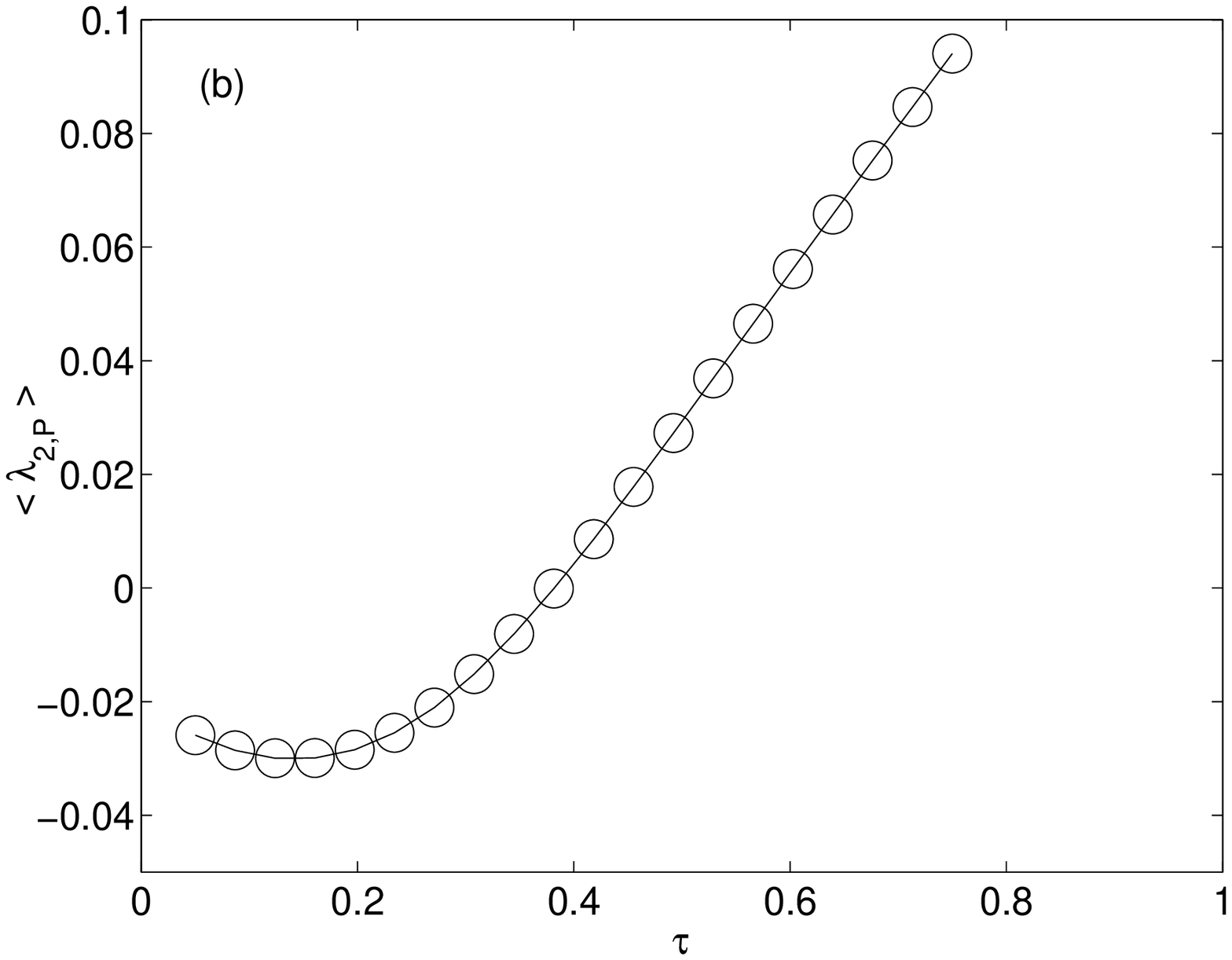}
\caption{\label{fig:meanpeval}(a) Plot of the mean eigenvalues 
$\langle\lambda_{i,P}\rangle$ of $P_{ij}$ as a function of the dimensionless 
time $\tau$.\\(b) Plot of $\langle\lambda_{2,P}\rangle$ as a function of the 
dimensionless time $\tau$, in the range $0<\tau\le0.75$.} 
\end{figure}

In Fig. \ref{fig:meanpeval}(a), we plot the mean eigenvalues 
$\langle\lambda_{i,P}\rangle$ as a function of the 
dimensionless time $\tau$, these evolve in a way that is similar to the 
temporal evolution of the kinetic-energy dissipation rate, 
with a peak in the magnitude of $\langle\lambda_{i,P}\rangle$, at cascade 
completion [cf. Fig. \ref{fig:prelim}(b)]. Similar results (not shown here) 
were obtained for the temporal evolution of $\langle\lambda_{i,Q}\rangle$. 
Ohkitani\cite{Ohkitani1}, in a $128^3$ DNS study of the unforced, 
incompressible, three-dimensional Euler equations, showed that 
$\lambda_{3,P}$ changes sign (locally, in 
regions of intense enstrophy) from positive to negative at early times. 
However, in our study, we find that $\langle\lambda_{2,P}\rangle$ changes 
sign [see Fig. \ref{fig:meanpeval}(b)] at $\tau=0.37$ from negative to 
positive, whereas $\langle\lambda_{3,P}\rangle$ remains negative at all 
times (not shown here). 

\begin{figure}
\includegraphics[height=1.9in]{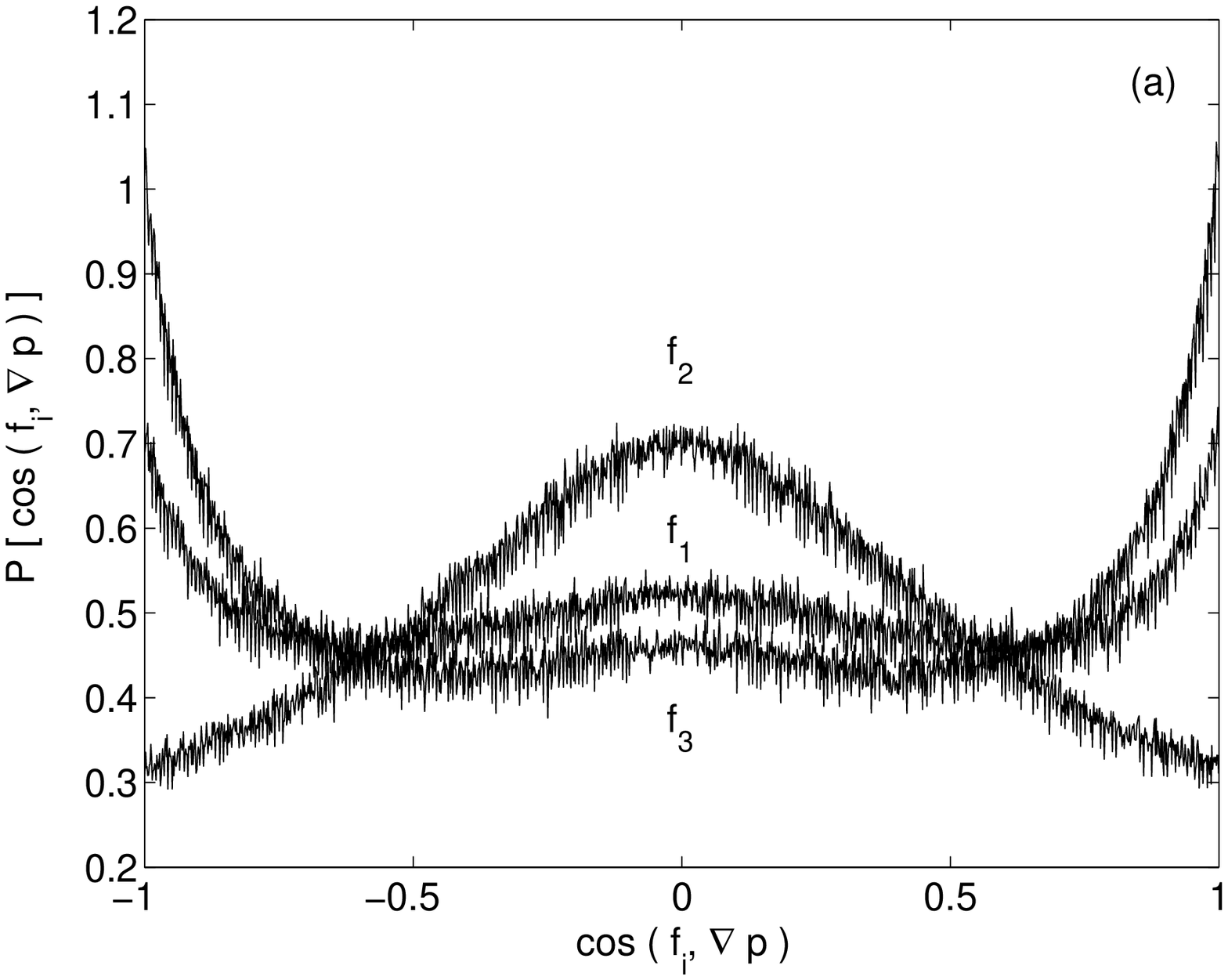}
\includegraphics[height=1.9in]{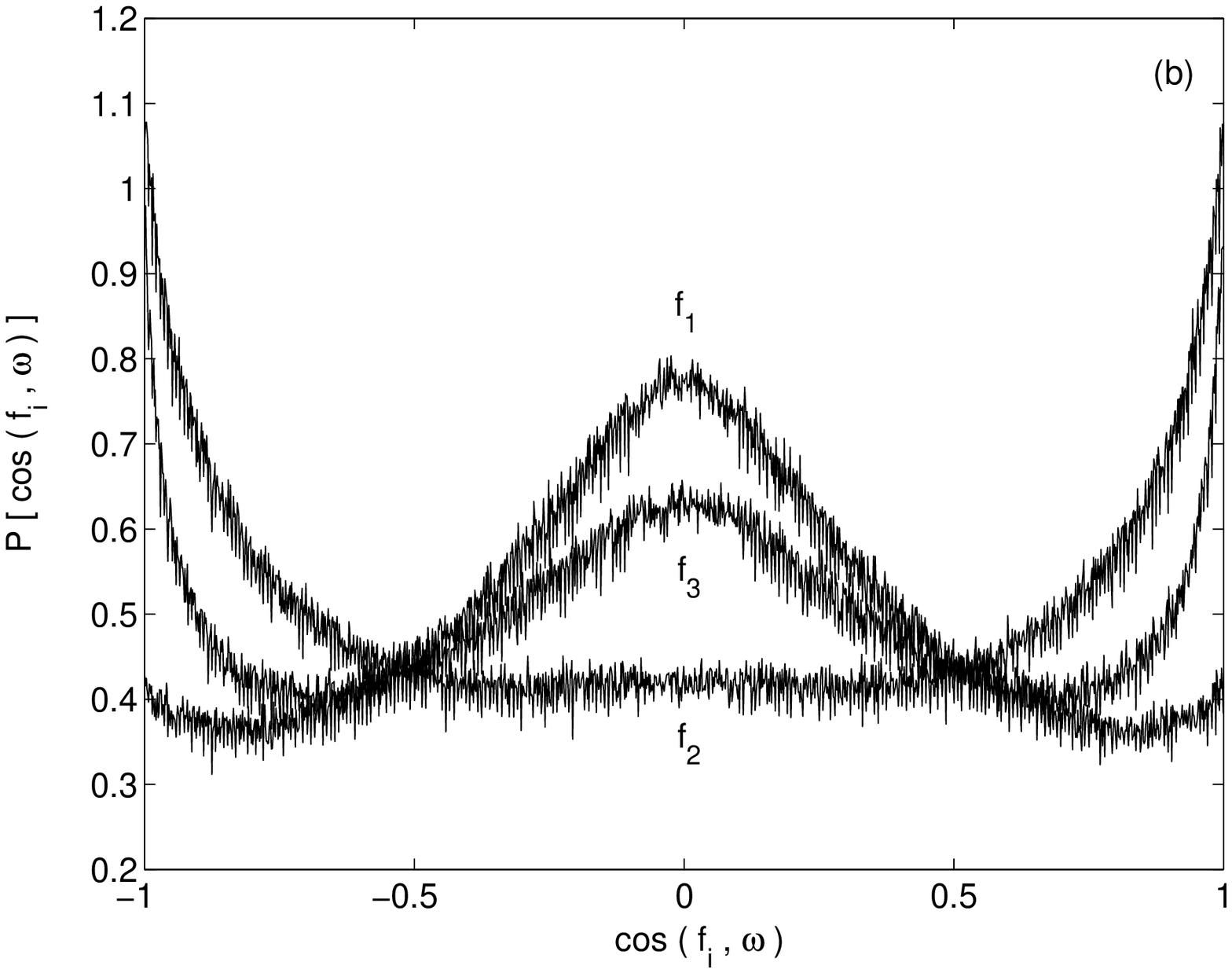}
\includegraphics[height=1.9in]{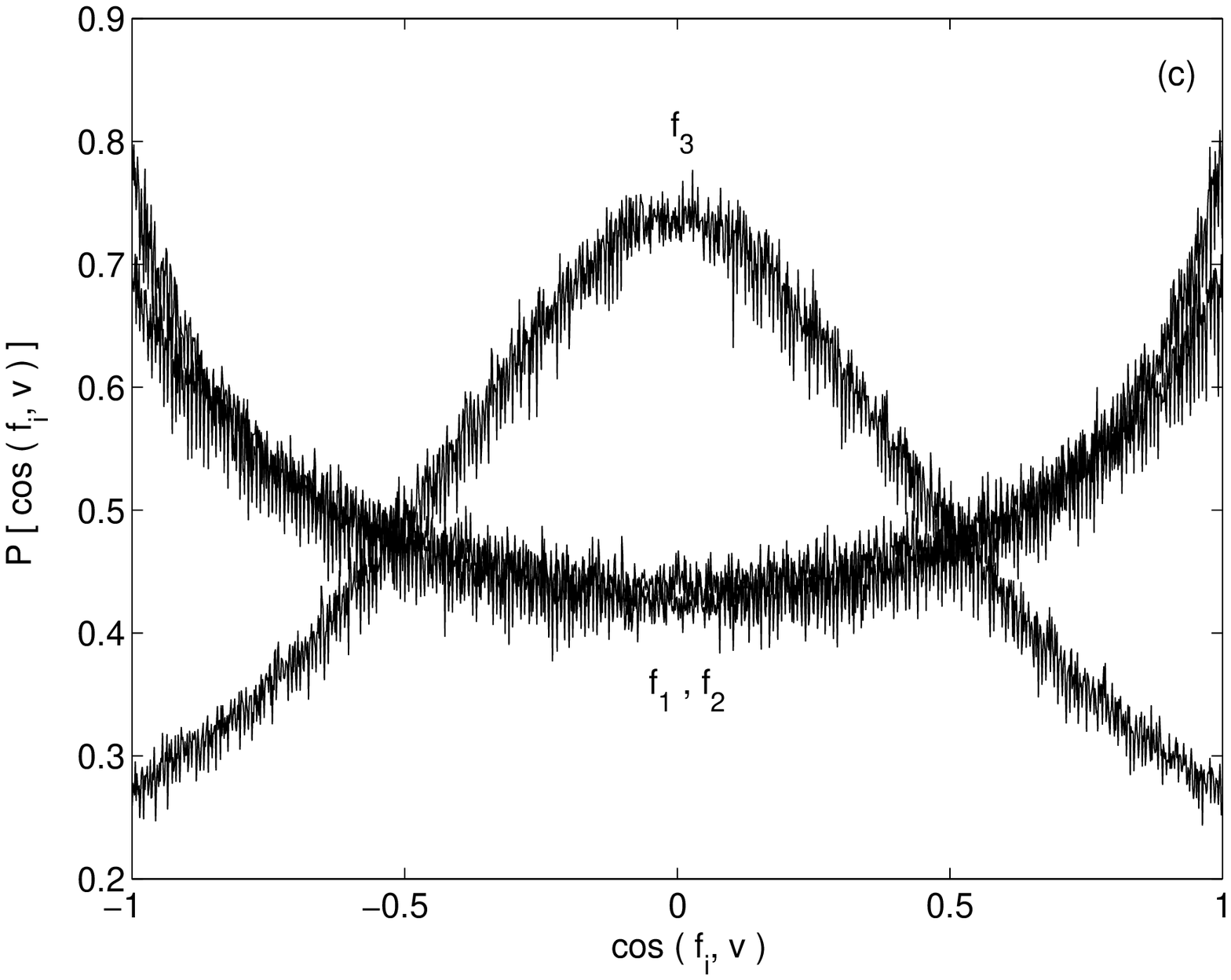}
\caption{\label{fig:anglefi}(a) Plot of the normalized probability 
distribution of the cosines of the angles between the eigenvectors 
$f_i$ of $P_{ij}$ and the pressure gradient $\nabla p$, at cascade completion.\\
(b) Plot of the normalized probability distribution of the cosines of 
the angles between the eigenvectors $f_i$ and the vorticity $\omega$, at 
cascade completion.\\(c) Plot of the normalized probability 
distribution of the cosines of the angles between the eigenvectors $f_i$ 
and the velocity ${\bf v}$, at cascade completion.}
\end{figure}

In Fig. \ref{fig:anglefi}(a), we plot the normalized probability distribution 
of the cosines of the angles between the eigenvectors $f_i$ of 
$P_{ij}$\cite{Fnmisc} and the pressure-gradient $\nabla p$ (see below), at 
cascade completion. In 
Fig. \ref{fig:anglefi}(b), we plot the normalized probability distribution 
of the cosines of the angles between the eigenvectors $f_i$ and $\omega$, at 
cascade completion. Ohkitani\cite{Ohkitani1} showed that $\omega$ is 
preferentially parallel (or anti-parallel) to the eigenvector $f_3$ 
corresponding to the pressure-hessian eigenvalue $\lambda_{3,P}$ smallest in 
magnitude, in contrast to our result, which shows that $\omega$ is 
preferentially parallel (or anti-parallel) with the eigenvector $f_2$ 
corresponding to the {\it intermediate} eigenvalue $\lambda_{2,P}$. In 
Fig. \ref{fig:anglefi}(c), we plot the normalized probability distribution 
of the cosines of the angles 
between the eigenvectors $f_i$ and the velocity ${\bf v}$, at 
cascade completion.
\subsection{Pressure Gradient}
\begin{figure}
\includegraphics[height=1.9in]{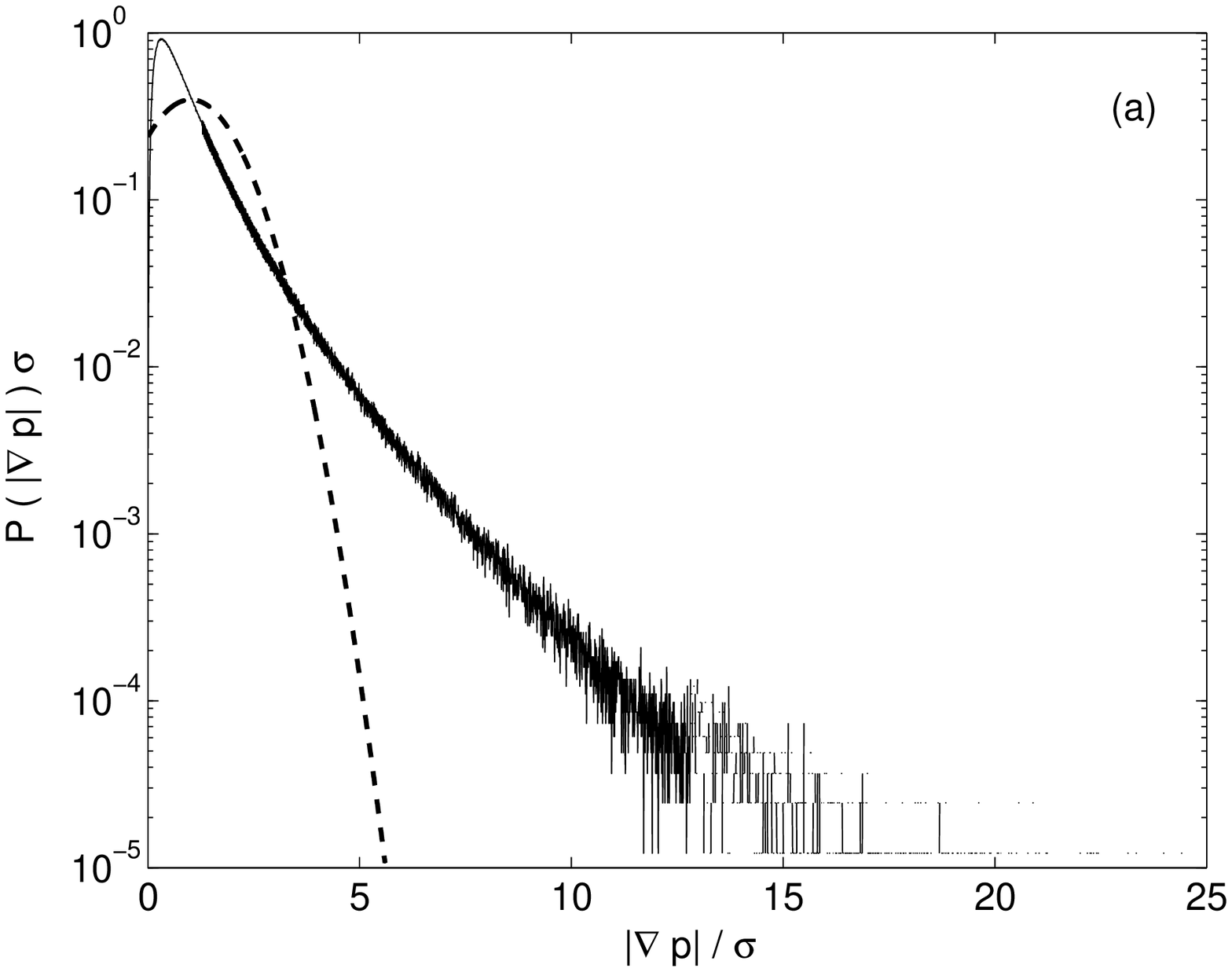}
\includegraphics[height=1.9in]{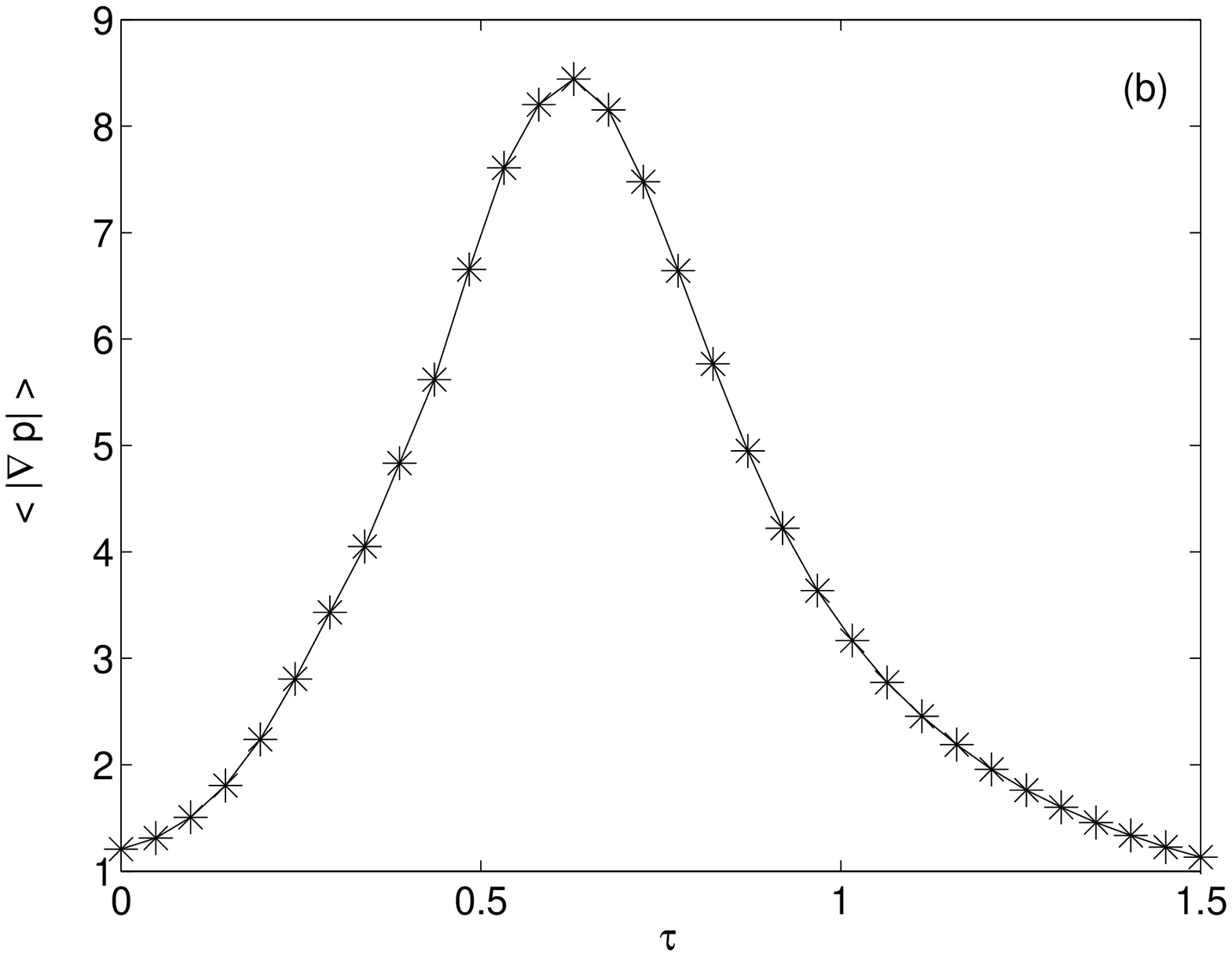}
\includegraphics[height=1.9in]{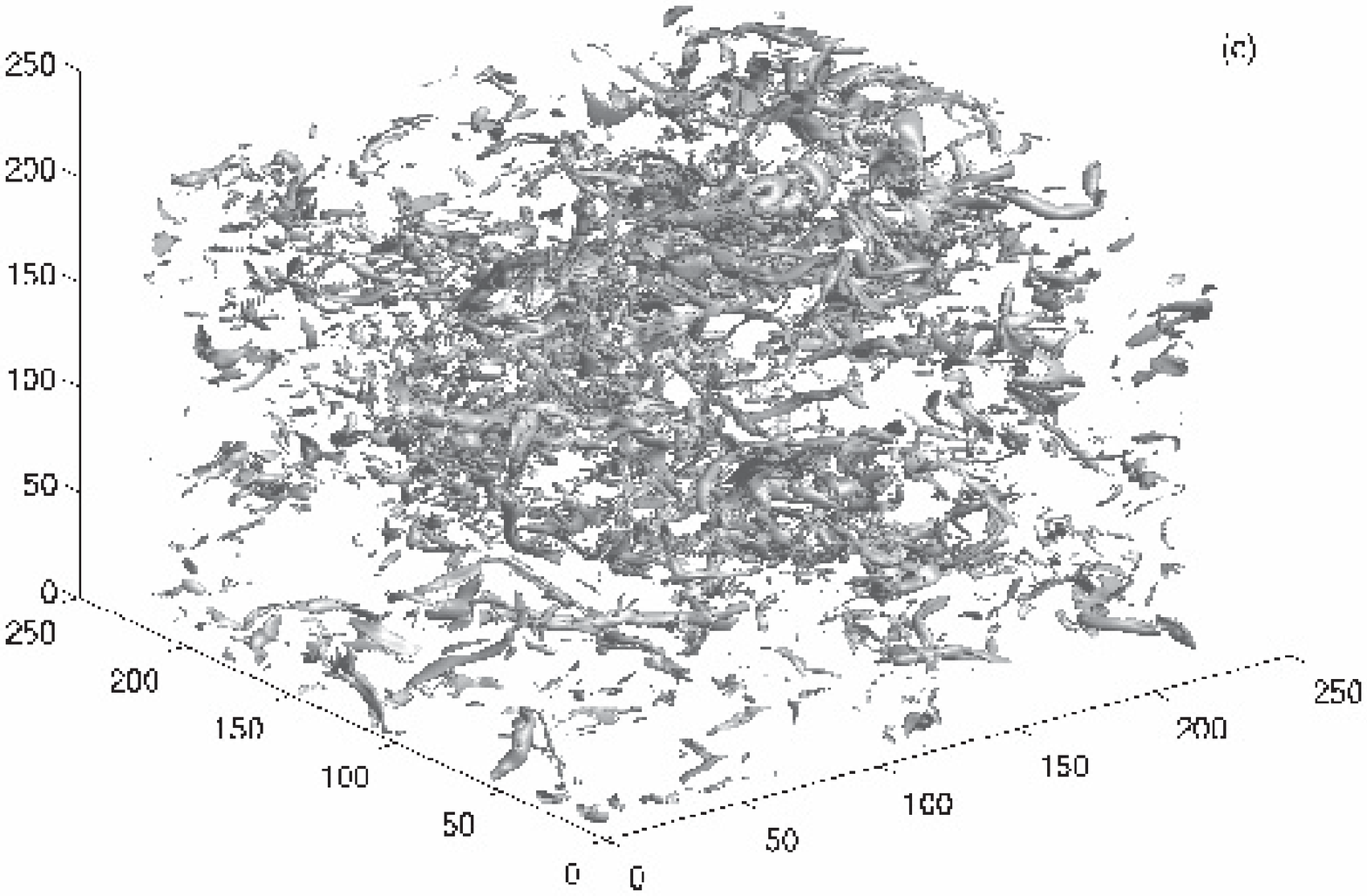}
\caption{\label{fig:pgrad}(a) Semilog plot of the normalized 
probability distribution ${\cal P}(|\nabla p|)$ ($|\nabla p|$ is the Euclidean 
norm of the pressure-gradient), at cascade completion. The 
dashed-line curve is a normalized Gaussian distribution for comparison.\\(b) 
Plot of the mean pressure-gradient norm $\langle|\nabla p|\rangle$ as a 
function of the dimensionless time $\tau$.\\(c) Plot of iso-$|\nabla p|$ 
surfaces for the 
isovalue $|\nabla p|=\langle|\nabla p|\rangle+2\sigma$, at cascade completion.} 
\end{figure}

A systematic numerical study of the pressure gradient is entirely lacking in 
both statistically steady and decaying turbulence. In 
Fig. \ref{fig:pgrad}(a), we plot the normalized probability 
distribution ${\cal P}(|\nabla p|)$ of the Euclidean norm 
($|{\bf x}|\equiv\sqrt{\sum_i x_i^2}$ for vector ${\bf x}$ with components 
$x_i$) of the pressure-gradient $\nabla p$ at cascade completion, which does 
not exhibit a stretched-exponential tail. The author has been unable to 
determine a functional form which gives a good fit for the tail of 
${\cal P}(|\nabla p|)$.

In Fig. \ref{fig:pgrad}(b), we plot the mean 
pressure-gradient norm $\langle|\nabla p|\rangle$ as a function of the 
dimensionless time $\tau$, which is observed to peak at $\tau=\tau_c$, as 
does the kinetic-energy dissipation rate [see Fig. \ref{fig:prelim}(b)].

In Fig. \ref{fig:pgrad}(c), we plot iso-$|\nabla p|$ surfaces for the isovalue 
$|\nabla p|=\langle|\nabla p|\rangle+2\sigma$, at cascade completion. The 
isosurfaces of intense pressure-gradient, which are filamentary in shape, are 
found to resemble the {\it low}-pressure 
isosurfaces in Fig. \ref{fig:isop}(b). Iso-$|\nabla p|$ surfaces in the 
range $|\nabla p|<(\langle|\nabla p|\rangle-\sigma)$ were not found to 
exhibit any particular structure at cascade completion.

\begin{figure}
\includegraphics[height=1.9in]{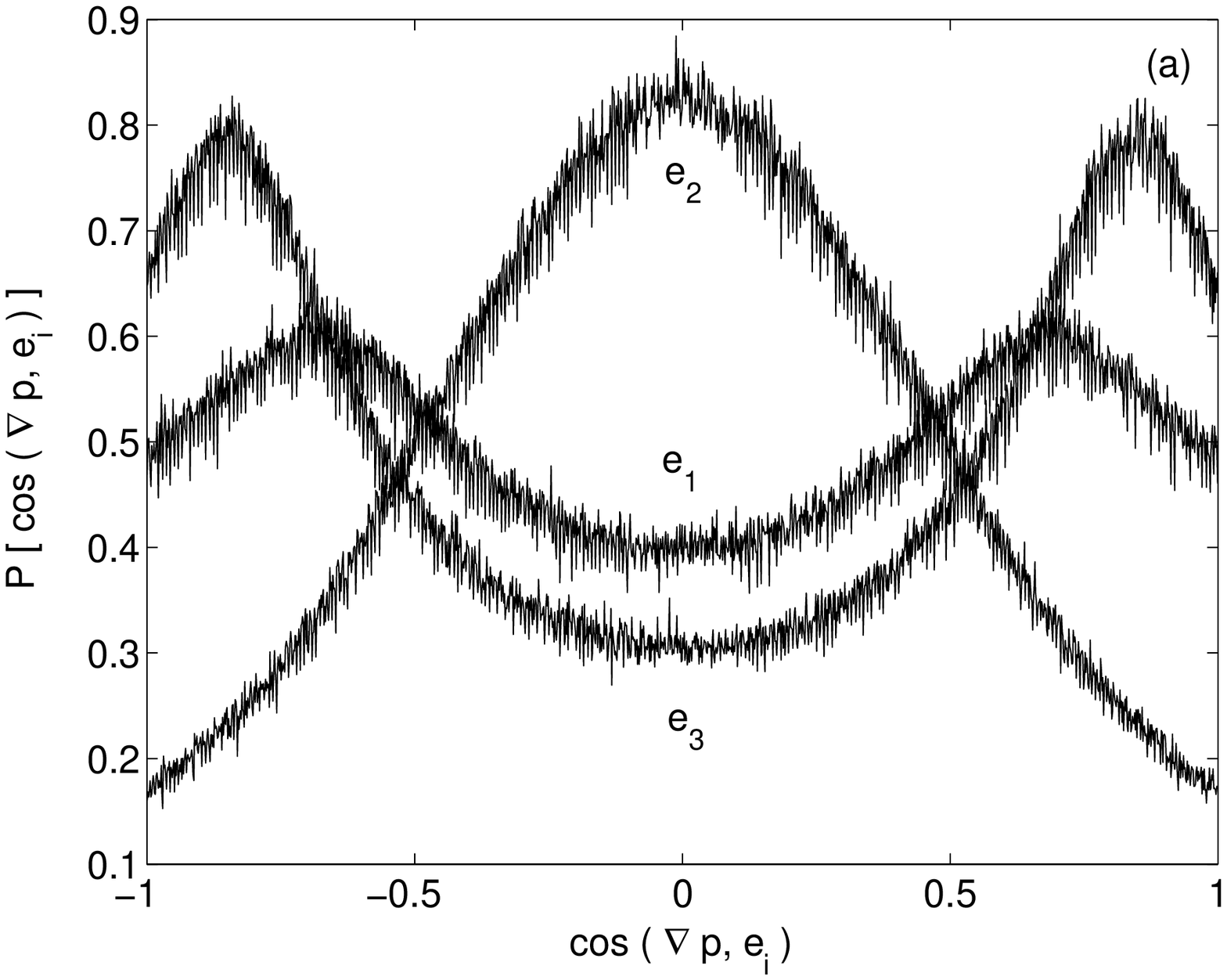}
\includegraphics[height=1.9in]{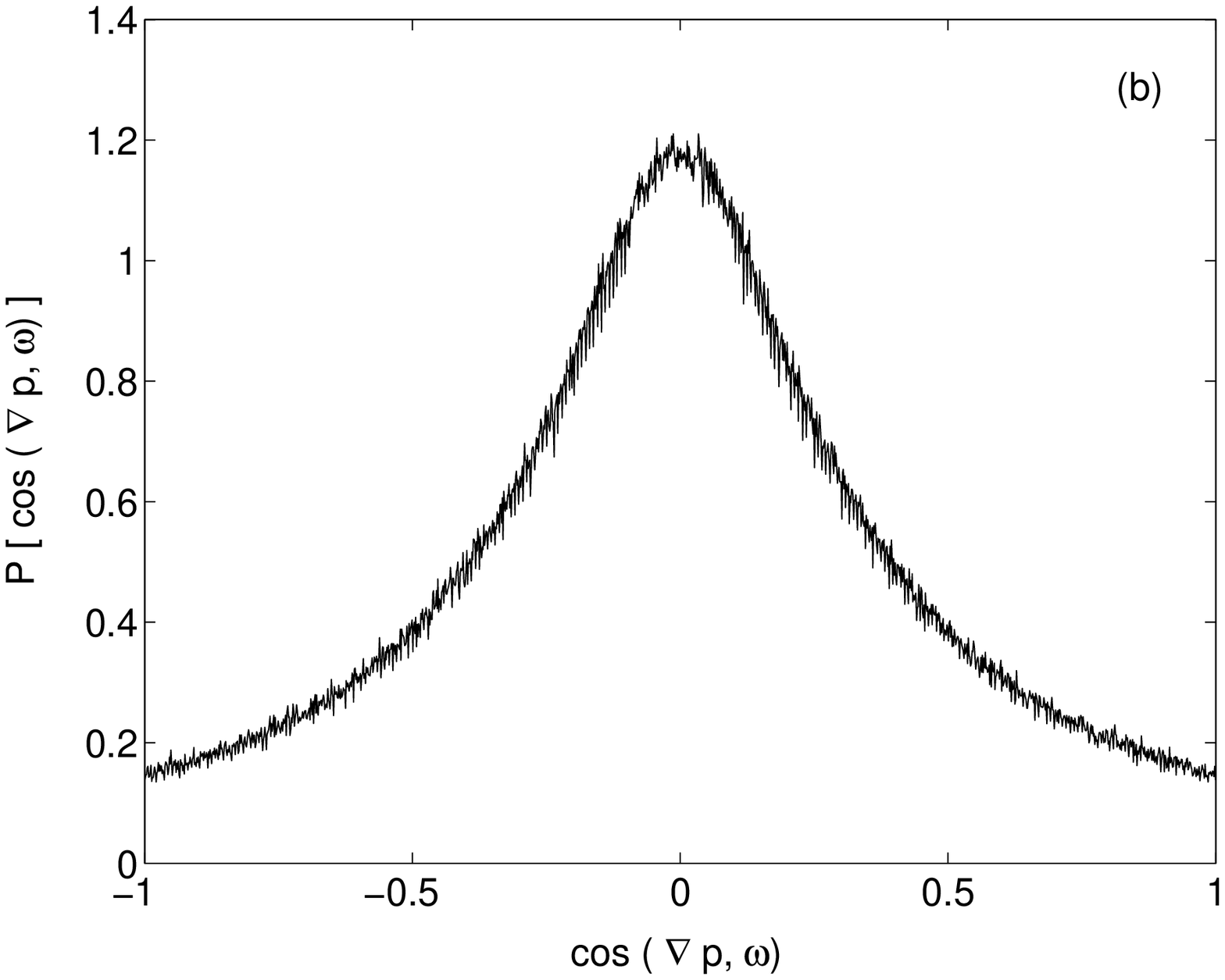}
\includegraphics[height=1.9in]{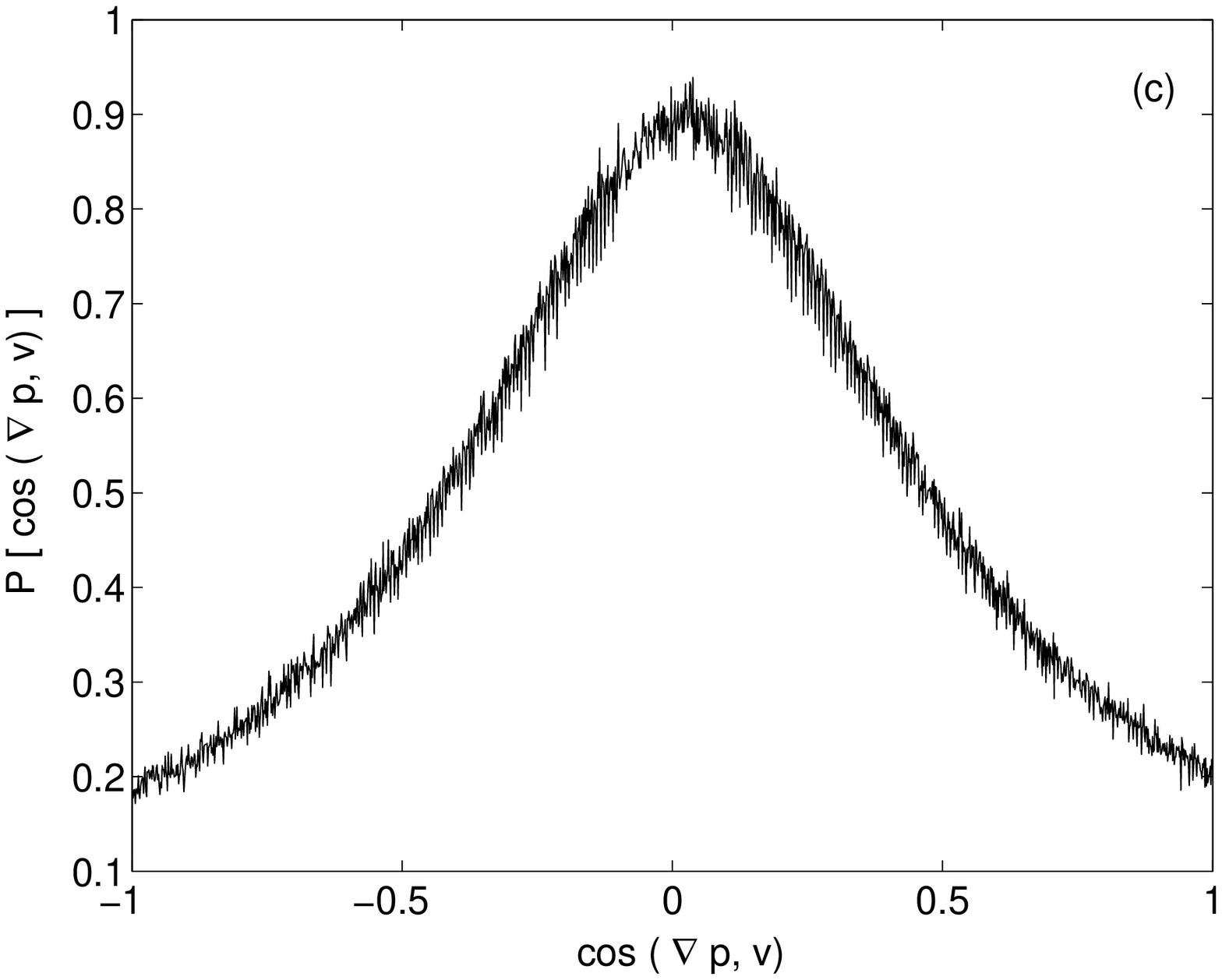}
\caption{\label{fig:anglepgrad}(a) Plot of the normalized probability 
distribution of the cosines of the angles between $\nabla p$ and the 
eigenvectors $e_i$ of the strain-rate tensor $S_{ij}$, at cascade 
completion.\\(b) Plot of the normalized probability distribution of 
the cosine of the angle between $\nabla p$ and $\omega$, at cascade 
completion.\\(c) Plot of the normalized probability distribution of 
the cosine of the angle between $\nabla p$ and ${\bf v}$, at cascade 
completion.}
\end{figure}

In Fig. \ref{fig:anglepgrad}(a), we plot the normalized probability 
distribution of the cosines of the angles between $\nabla p$ and the 
eigenvectors $e_i$ of $S_{ij}$. Ashurst, {\it et.al.}\cite{Ashurst}, noted a 
tendency for the alignment of $\nabla p$ (caused by velocity fluctuations 
alone) with the most compressive strain direction $e_3$ in a $128^3$ DNS 
study of a statistically steady turbulent shear-flow. However, in our 
study, $\nabla p$ is not found to be preferentially parallel (or 
anti-parallel) with $e_3$, and we observe a 
peak in the magnitude of ${\cal P}[\cos(\nabla p, e_3)]$ at 
$|\cos(\nabla p,e_3)|\approx0.86$, which indicates a preferential relative 
angle $\approx\pi/6$. In Fig. \ref{fig:anglepgrad}(b), we plot the 
normalized probability distribution of the cosine of the angle between 
$\nabla p$ and $\omega$, at cascade completion. In Fig. 
\ref{fig:anglepgrad}(c), we plot the normalized probability distribution of 
the cosine of the angle between $\nabla p$ and ${\bf v}$, at cascade 
completion. Both $\omega$ and ${\bf v}$ are observed to be preferentially 
perpendicular to $\nabla p$, in agreement with corresponding results from a 
$128^3$ DNS study of the unforced, incompressible, three-dimensional Euler 
equations, due to Ohkitani and Kishiba\cite{Ohkitani2}. 
\subsection{General Alignment Picture}
\begin{figure}
\includegraphics[height=1.9in]{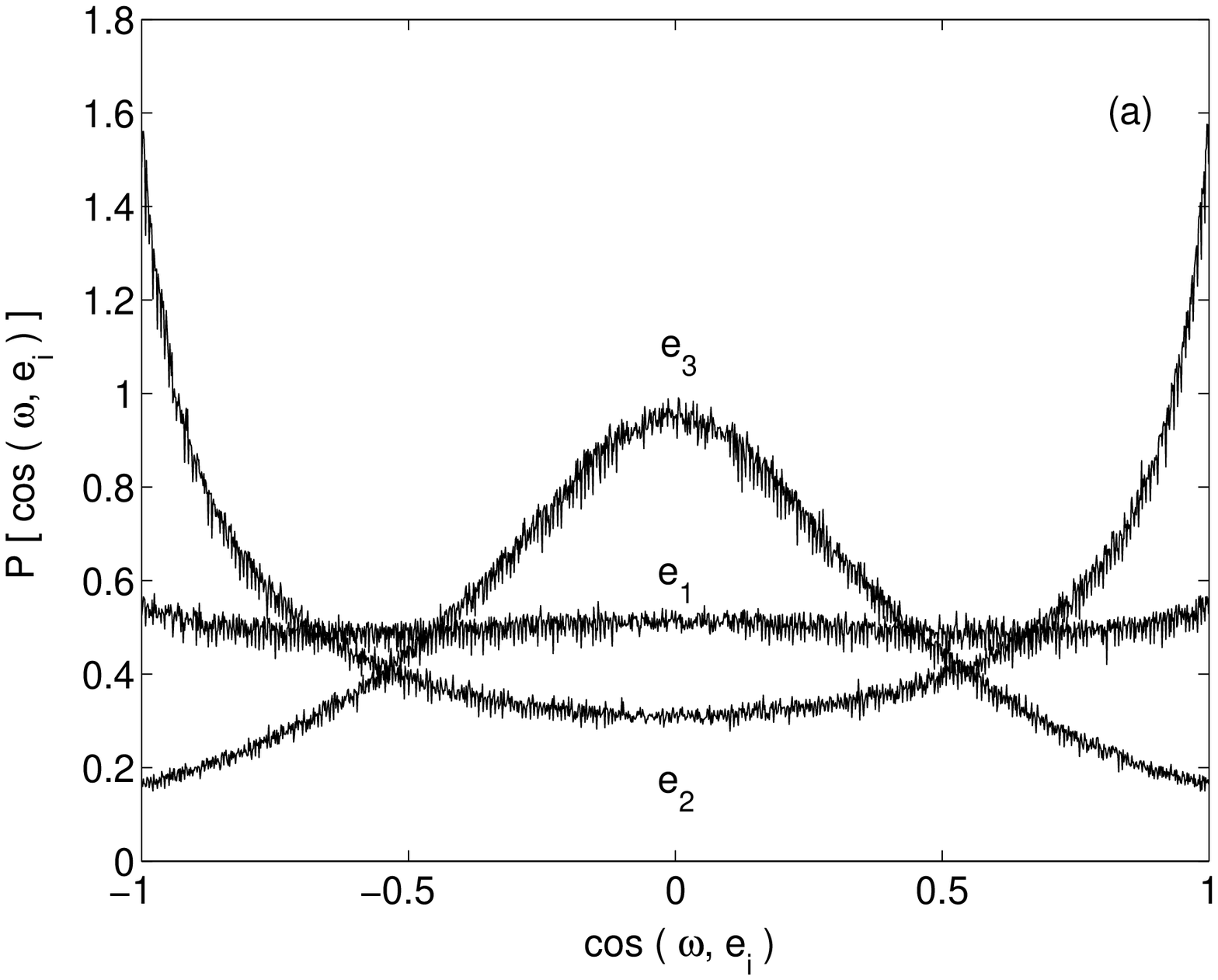}
\includegraphics[height=1.9in]{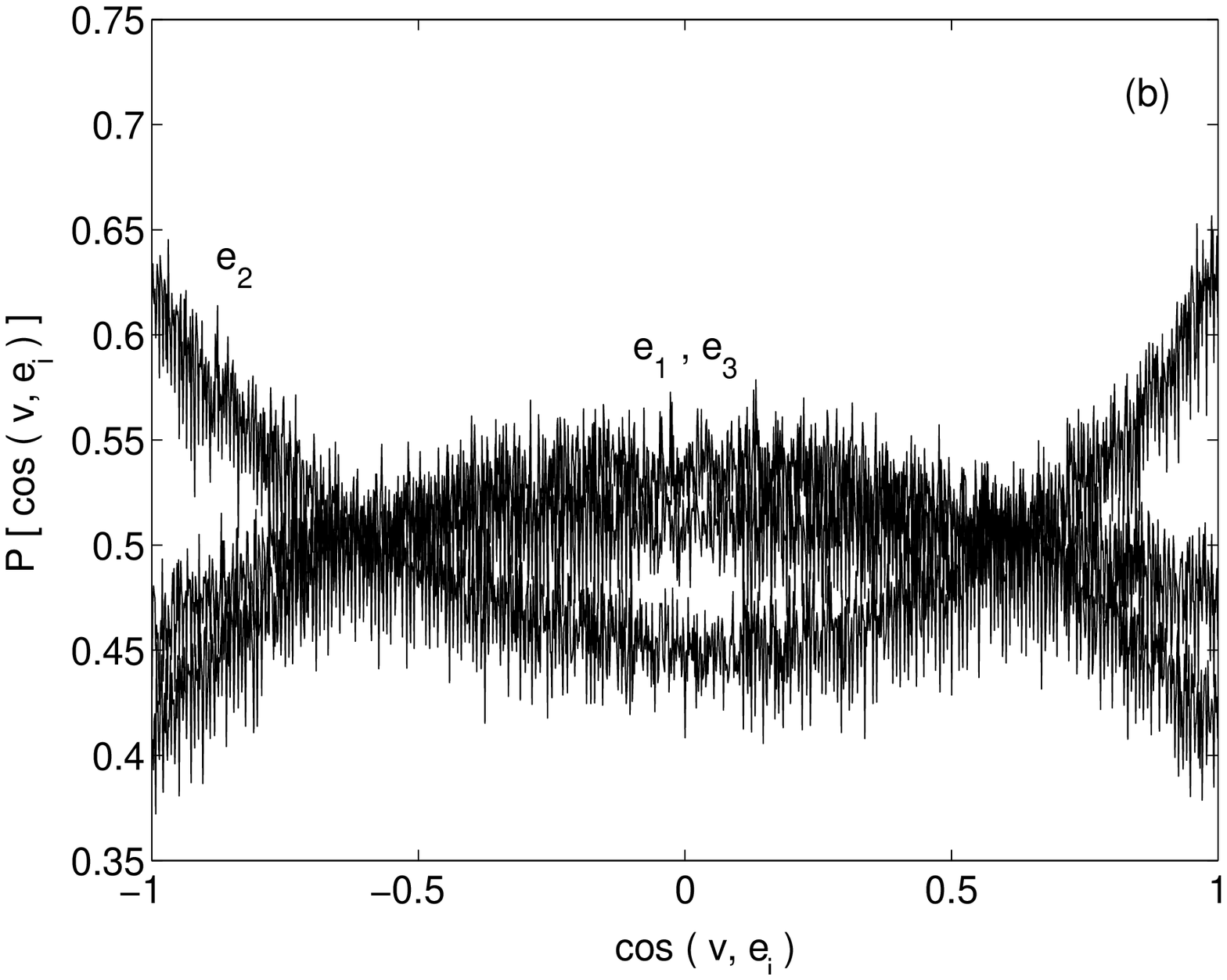}
\includegraphics[height=1.9in]{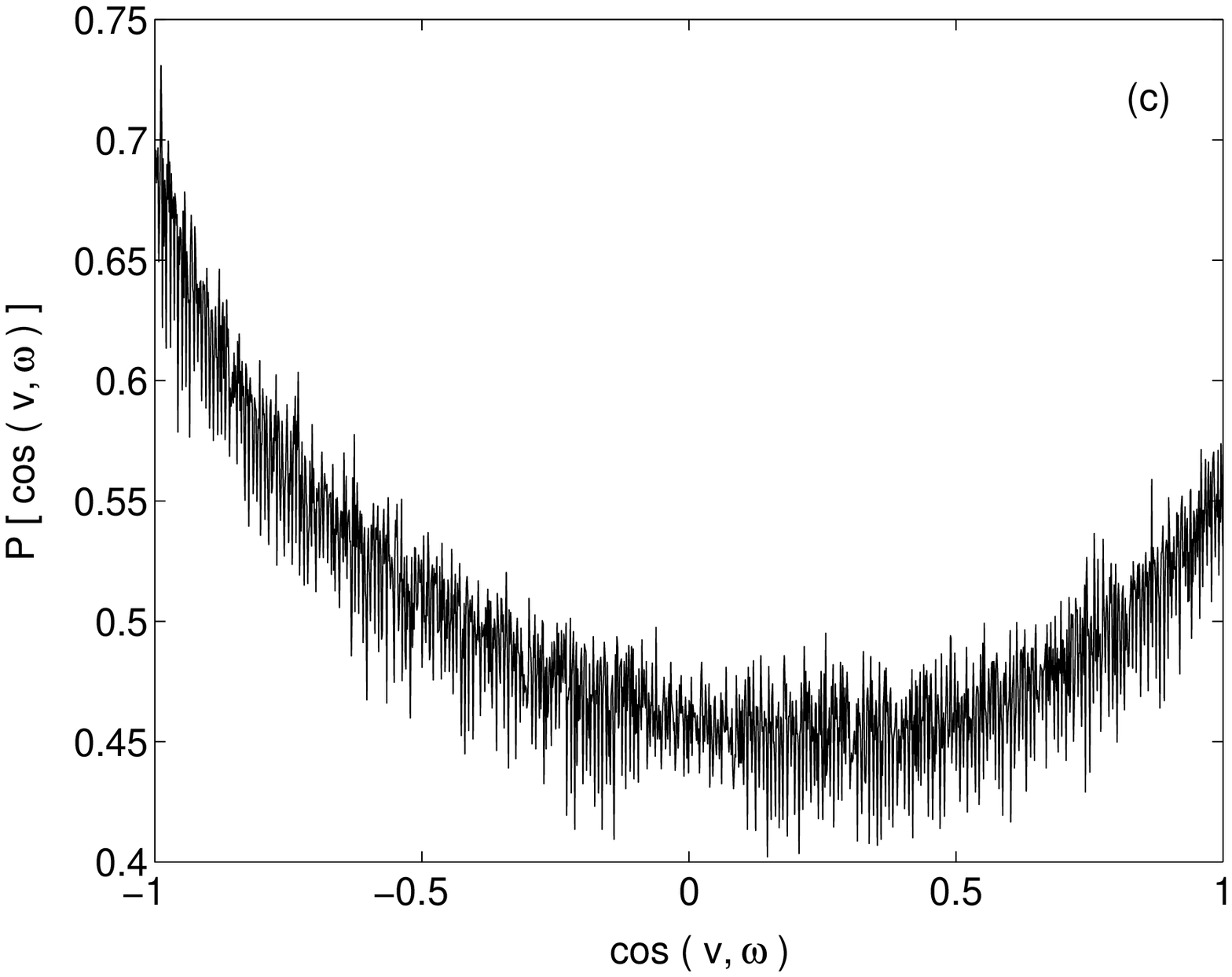}
\caption{\label{fig:misc}(a) Plot of the normalized probability 
distribution of the cosines of the angles between $\omega$ and the 
eigenvectors $e_i$, at cascade completion.\\
(b) Plot of the normalized probability distribution of the cosines of 
the angles between ${\bf v}$ and the eigenvectors $e_i$, at cascade 
completion.\\(c) Plot of the normalized probability distribution of 
the cosine of the angle between ${\bf v}$ and $\omega$, at cascade completion.}
\end{figure}

\begin{figure}
\includegraphics[height=1.9in]{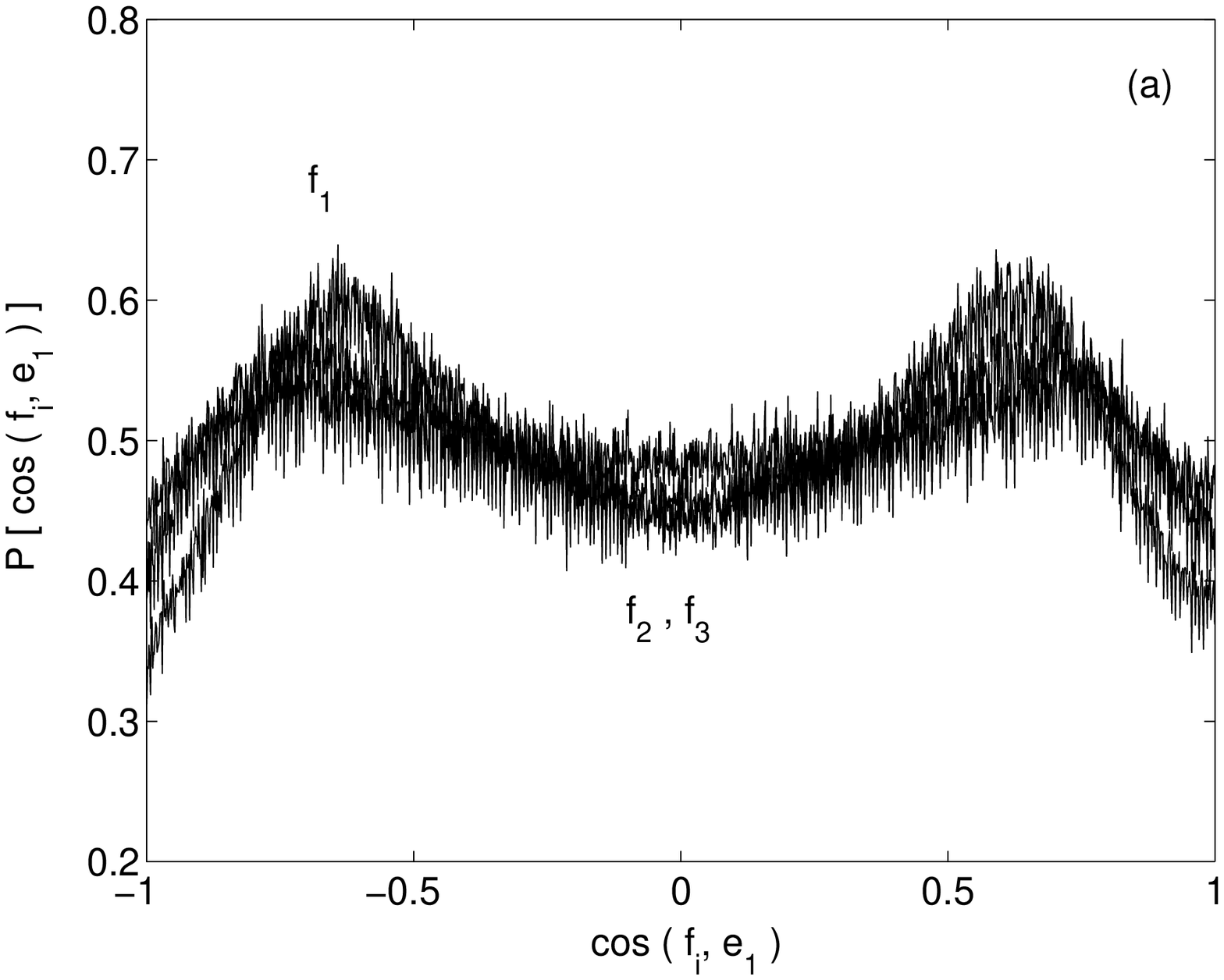}
\includegraphics[height=1.9in]{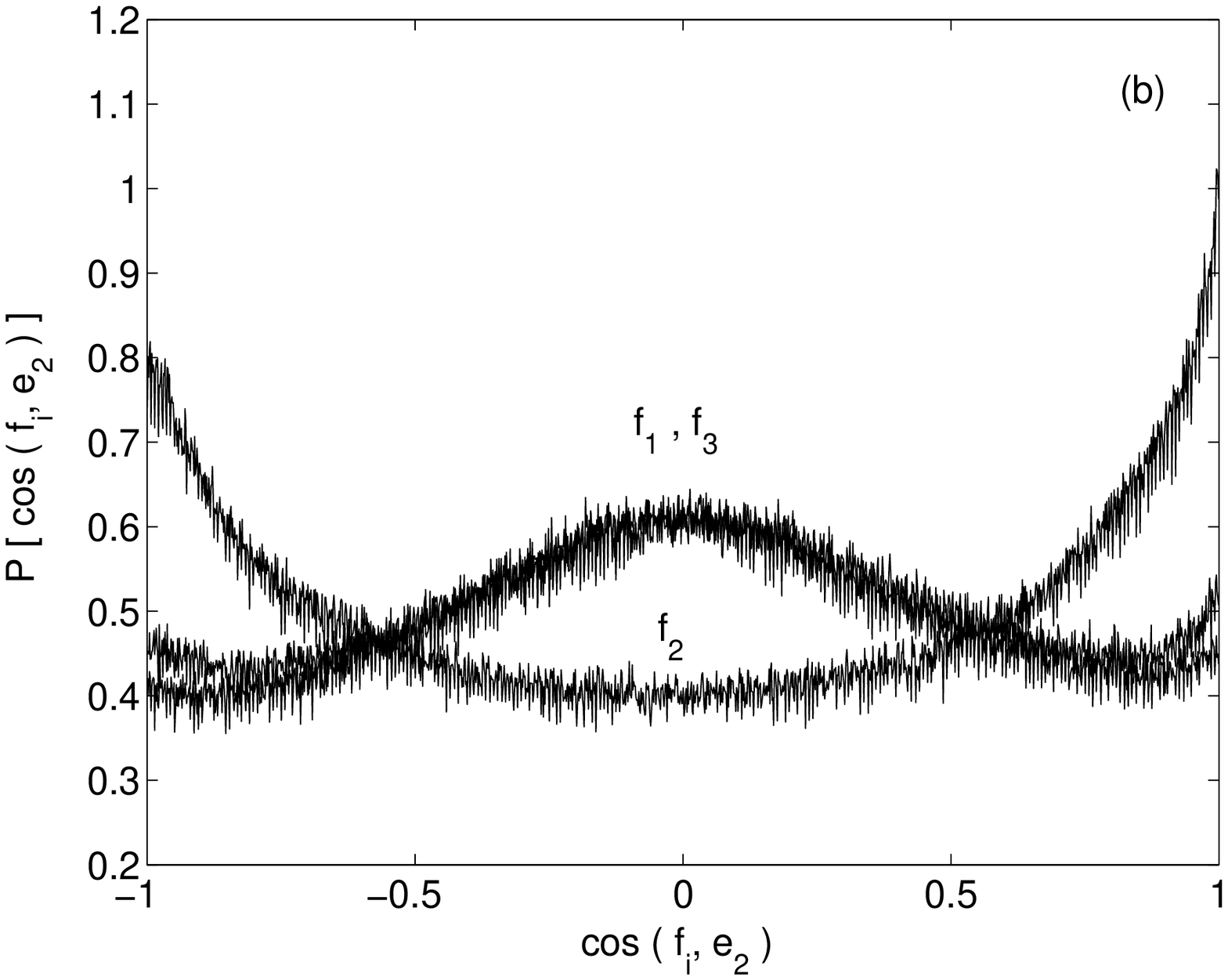}
\includegraphics[height=1.9in]{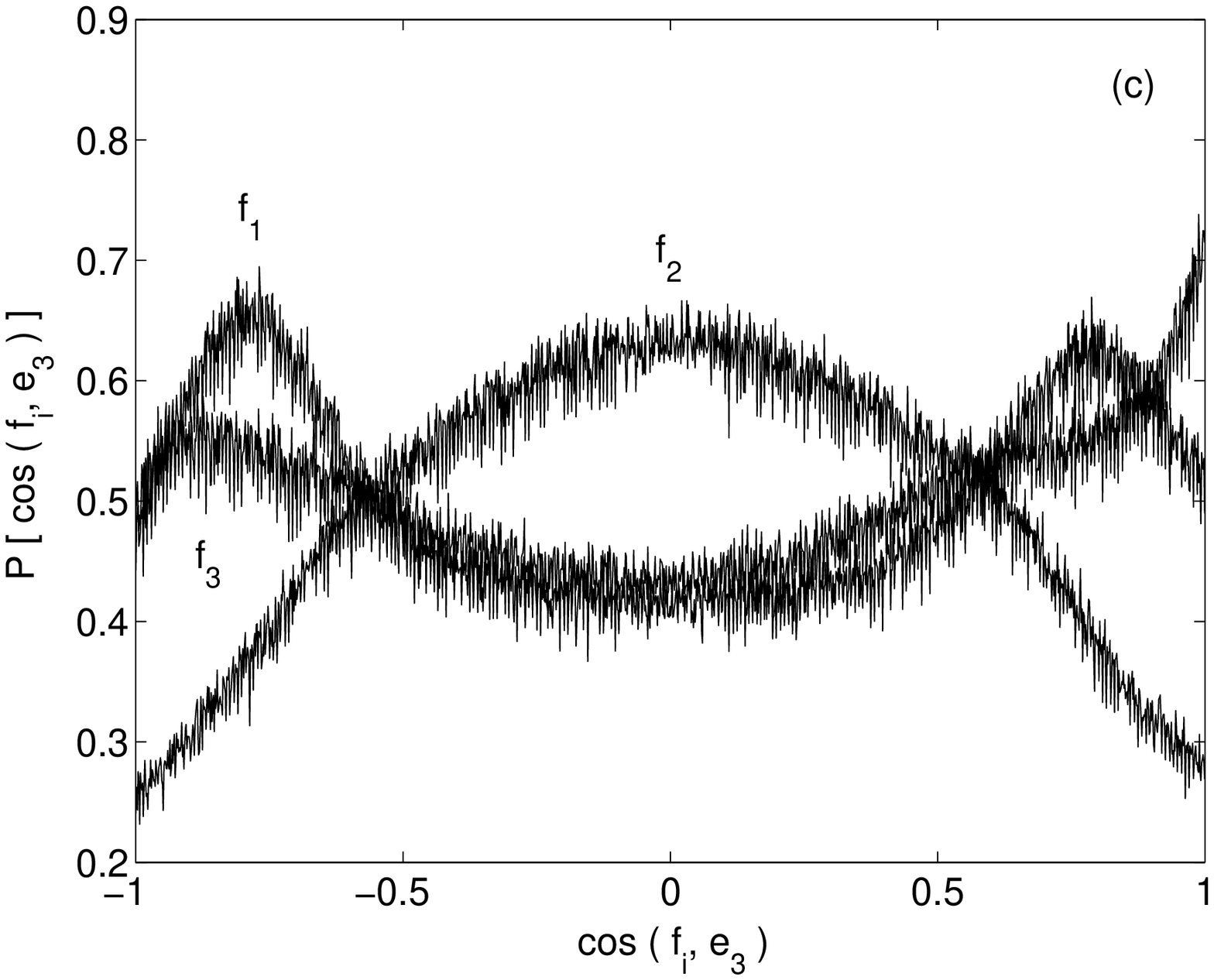}
\caption{\label{fig:rel}(a) Plot of the normalized probability 
distribution of the cosines of the angles between the eigenvectors $f_i$ and 
$e_1$, at cascade completion.\\(b) Plot of the normalized 
probability distribution of the cosines of the angles between the eigenvectors 
$f_i$ and $e_2$, at cascade completion.\\(c) Plot of the 
normalized probability distribution of the cosines of the angles between the 
eigenvectors $f_i$ and $e_3$, at cascade completion.}
\end{figure}

Ohkitani and Kishiba\cite{Ohkitani2} have found the orientations (locally, in 
regions of intense enstrophy) amongst the set of vectors $(e_i, f_i, \omega)$, 
excluding the pressure gradient and the velocity. In Figs. \ref{fig:anglefi} 
and \ref{fig:anglepgrad}, we have plotted possible 
alignments between the sets of vectors ($f_i$, $\nabla p$) and 
($e_i$, ${\bf v}$, $\omega$) and for completeness, we plot in Figs. 
\ref{fig:misc} and \ref{fig:rel}, remaining alignments amongst the set 
of vectors ($e_i$, ${\bf v}$,  $\omega$) and amongst the eigenvector 
bases ($f_i$, $e_i$), at cascade completion. 

As is well-known, in both decaying\cite{Kalelkar} and statistically 
steady\cite{Vincent} turbulence, there is an increased probability for 
alignment (or anti-alignment) of the intermediate strain-rate 
eigenvector $e_2$ with $\omega$, relative to the alignments between 
$e_1$ and $e_3$ with $\omega$. In Fig. \ref{fig:misc}(a), we plot the 
normalized probability distribution of the cosines of the angles 
between $\omega$ and $e_i$ at cascade completion, which reaffirms this 
result for decaying turbulence. In Fig. \ref{fig:misc}(b), we plot 
the normalized probability distribution of the cosines of the angles 
between ${\bf v}$ and $e_i$, at cascade completion. In Fig. \ref{fig:misc}(c), 
we plot the normalized 
probability distribution of the cosine of the angle between ${\bf v}$ 
and $\omega$, at cascade completion. It is interesting to note that the 
alignment plots in Figs. \ref{fig:anglefi}, \ref{fig:anglepgrad}, and 
\ref{fig:misc}(a),(b) are roughly symmetrically placed about $\cos\theta=0$, 
however Fig. \ref{fig:misc}(c) is distinctly asymmetric, with a greater 
probability for ${\bf v}$ and $\omega$ to be 
{\it anti}-parallel. This asymmetry has been noted in a laboratory 
experiment\cite{Kit} of decaying turbulent flow past a grid, and is 
plausibly due to effects of the kinetic helicity\cite{Betchov} (an 
invariant of the Euler equations) on the decay process. 

In Fig. \ref{fig:rel}(a), ${\cal P}[\cos(f_1,e_1)]$ is found to peak at 
$|\cos(f_1,e_1)|\approx0.71$, which indicates a preferential relative 
angle $\approx\pi/4$, in agreement with corresponding results due to 
Ohkitani and Kishiba\cite{Ohkitani2}. The only 
distinct feature in Figs. \ref{fig:rel}(b),(c) is that $f_2$ is preferentially 
parallel (or anti-parallel) to $e_2$ (and perpendicular to $e_3$).

\begin{figure}
\includegraphics[height=1.9in]{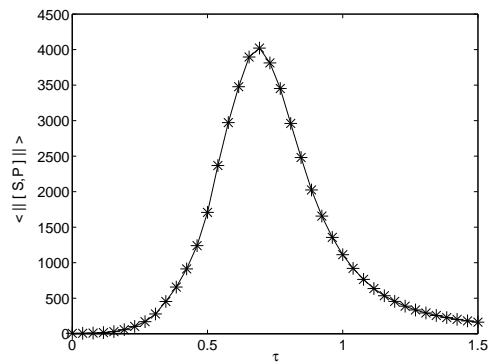}
\caption{\label{fig:norm}Plot of the mean norm $\langle||[S,P]||\rangle$ (see 
the text for the definition of the norm) of the commutator $[S,P]=SP-PS$ of 
the strain-rate $S$ and the pressure-hessian $P$ tensors, as a function of 
the dimensionless time $\tau$.}
\end{figure}

Ohkitani\cite{Ohkitani1} has conjectured that the pressure-hessian tensor $P$ 
(with components $P_{ij}$) and the strain-rate tensor $S$ (with 
components $S_{ij}$) in general are {\it not} commutative, and therefore 
cannot be simultaneously diagonalized. It is of interest to determine 
the (statistically preferred) relative orientation, between the two frames 
with respect to which $S$ and $P$ are diagonalized. Ohkitani and 
Kishiba\cite{Ohkitani2} have conjectured that the configuration of relative 
alignments between the eigenvector bases ($f_i,e_i$) is one with ``least 
commutativity between $S$ and $P$ out of all possibilities with one axis 
in common" (p. $414$, Ref. \cite{Ohkitani2}) at cascade completion. In 
order to test these conjectures, we choose the standard matrix norm 
$||A||=($maximum eigenvalue of $A^TA)^{1/2}$\cite{Strang}, where 
the superscript $T$ denotes the transpose-conjugate. In Fig. \ref{fig:norm}, 
we plot the mean norm $\langle||[S,P]||\rangle$ of the 
commutator $[S,P]=SP-PS$\cite{Fncom} as a function of the dimensionless 
time $\tau$ and find that the value peaks at $\tau=0.71$. The 
{\it value} of the mean norm 
depends on the choice of the norm; however, the trend (which is independent 
of this choice) indicates that the relative configuration of the 
eigenvector bases $(f_i,e_i)$ is ``least commutative" at $\tau=0.71$, which is 
equal to the time at which the kinetic energy-dissipation rate peaks, in 
accord with the Ohkitani and Kishiba\cite{Ohkitani2} conjecture. 
\section{Conclusion}
To summarize, we have presented results from a systematic numerical study of
pressure fluctuations in an unforced, incompressible, homogeneous, and 
isotropic turbulent fluid. At cascade completion, isosurfaces of low pressure 
are found to be organised as slender filaments, whereas the predominant 
pressure isostructures appear sheet-like. We have exhibited several new 
results, including plots of the probability distributions of the spatial 
pressure-difference, the pressure-gradient norm, and the eigenvalues of the 
pressure-hessian tensor, at cascade completion. Plots of the temporal 
evolution of the mean pressure-gradient norm, and the mean eigenvalues 
of the pressure-hessian tensor have also been exhibited. We have found the 
statistically preferred orientations between the eigenvectors of the 
pressure-hessian tensor, the pressure-gradient, the eigenvectors of the 
strain-rate tensor, the 
vorticity, and the velocity at cascade completion. Statistical properties of 
the non-local part of the pressure-hessian tensor have also been exhibited, for 
the first time. We have presented numerical 
tests (in the viscous case) of some conjectures for the unforced, 
incompressible, three-dimensional Euler equations, proposed in earlier studies.
\section{Acknowledgement}
The author thanks T. Kalelkar and R. Pandit for discussions, D. Mitra for 
the code, SERC (IISc) for computational resources, and CSIR (India) for 
financial support.


\newpage 
\begin{thebibliography}{99}
\addcontentsline{toc}{chapter}{Bibliography}
\bibitem{Pumir} A. Pumir, Phys. Fluids {\bf 6}, 2071 (1994).
\bibitem{Cao} N. Cao, S. Chen, and G. Doolen, Phys. Fluids {\bf 11}, 2235 
(1999).
\bibitem{Gotoh} T. Gotoh and D. Fukayama, Phys. Rev. Lett. {\bf 86}, 3775 
(2001).
\bibitem{Ishihara} T. Ishihara, Y. Kaneda, M. Yokokawa, K. Itakura, and 
A. Uno, J. Phys. Soc. Jpn. {\bf 72}, 983 (2003). 
\bibitem{Ashurst} W. Ashurst, J. Chen, and M. Rogers, Phys. Fluids {\bf 30}, 
3293 (1987).
\bibitem{She} Z. She, E. Jackson, and S. Orszag, Nature, {\bf 344}, 226 
(1990); Z. She, E. Jackson, and S. Orszag, Proc. Roy. Soc. London, Ser. A, 
{\bf 434}, 101 (1991).
\bibitem{Fauve} S. Fauve, C. Laroche, and B. Castaing, J. Phys. II (France) 
{\bf 3}, 271 (1993).
\bibitem{Abry} P. Abry, S. Fauve, P. Flandrin, and C. Laroche, J. Phys. II 
(France) {\bf 4}, 725 (1994).
\bibitem{Douady} S. Douady, Y. Couder, and M. Brachet, Phys. Rev. Lett. 
{\bf 67}, 983 (1991).
\bibitem{Villermaux} E. Villermaux, B. Sixou, and Y. Gagne, Phys. Fluids 
{\bf 7}, 2008 (1995).
\bibitem{Monin} A. Monin and A. Yaglom, {\it Statistical Fluid 
Mechanics}, edited by J. Lumley (MIT Press, Cambridge, 1975), Vol. {\bf 2}.
\bibitem{Schumann} U. Schumann and G. Patterson, J. Fluid Mech. {\bf 88}, 
685 (1978). 
\bibitem{Ohkitani1} K. Ohkitani, Phys. Fluids A {\bf 5}, 2570 (1993).
\bibitem{Ohkitani2} K. Ohkitani and S. Kishiba, Phys. Fluids {\bf 7}, 411 
(1995).
\bibitem{Dhar} S. Dhar, A. Sain, and R. Pandit, Phys. Rev. Lett. {\bf 78}, 
2964 (1997).
\bibitem{Borue} V. Borue and S. Orszag, Phys. Rev. E, {\bf 51}, R856 (1995).
\bibitem{Yamamoto} K. Yamamoto and I. Hosokawa, J. Phys. Soc. Jpn. 
{\bf 57}, 1532 (1988).
\bibitem{Kalelkar} C. Kalelkar, Phys. Rev. E {\bf 72}, 056307 (2005).
\bibitem{Tsuji} Y. Tsuji and T. Ishihara, Phys. Rev. E {\bf 68}, 026309 (2003).
\bibitem{Nelkin} M. Nelkin, Adv. Phys. {\bf 43}, 143 (1994).
\bibitem{Vincent} A. Vincent and M. Meneguzzi, J. Fluid Mech. {\bf 225}, 1 
(1991).
\bibitem{Holzer} M. Holzer and E. Siggia, Phys. Fluids A {\bf 5}, 2525 (1993).
\bibitem{Fng} As is well-known (see Ref. \cite{Monin} for instance), a 
Gaussian velocity probability distribution implies {\it zero} odd-order 
velocity correlations, in contradiction with numerical and laboratory 
results. 
\bibitem{Brachet} M. Brachet, Fluid Dyn. Res. {\bf 8}, 1 (1991).
\bibitem{Taylor} G. Taylor and A. Green, Proc. Roy. Soc. London, Ser. A 
{\bf 158}, 499 (1937).
\bibitem{Fneig} In Ref. \cite{Kalelkar}, the statistically preferred ratio 
of the mean strain-rate eigenvalues $\langle\lambda_{1,S}\rangle:\langle\lambda_{2,S}\rangle:\langle\lambda_{3,S}\rangle$ was found to equal $5:1:-6$, at 
cascade completion.
\bibitem{Fnloc} From Ref. \cite{Ohkitani2}, 
$Q_{ij}({\bf x})\equiv\oint q_{ij}({\bf x}-{\bf y})\nabla^2_yp({\bf y})
d{\bf y}$, $q_{ij}({\bf x})=(|{\bf x}|^2\delta_{ij}-3x_ix_j)/
(4\pi|{\bf x}|^5)$, $\nabla^2_y=\partial^2/\partial y_i\partial y_i$, where 
the integral is taken in the sense of principal values. 
\bibitem{Fnmisc} The alignment results for the eigenvectors of $Q_{ij}$ were 
found to resemble those for $f_i$.
\bibitem{Kit} E. Kit, A. Tsinober, J. Balint, J. Wallace, and E. Levich, Phys. 
Fluids {\bf 30}, 3323 (1987).
\bibitem{Betchov} R. Betchov, Phys. Fluids {\bf 4}, 925 (1961).
\bibitem{Strang} G. Strang, {\it Linear Algebra and its Applications}, 
(Thomson Learning, London, 1988).
\bibitem{Fncom} The corresponding plot for $[S,Q]$ is redundant since 
$[S,Q]=[S,P]$.
\end{thebibliography}
\end{document}